\documentclass[10pt]{iopart}
\usepackage[a4paper,margin=1.85cm]{geometry}  

\usepackage{tgtermes}
\usepackage[T1]{fontenc}

\usepackage{graphicx}  

\expandafter\let\csname equation*\endcsname\relax
\expandafter\let\csname endequation*\endcsname\relax
\usepackage{amsmath}
\usepackage{amssymb}  

\usepackage[frozencache]{minted}
\usepackage{overpic}

\providecommand{\todo}[1]{}
\newcommand{\dens}[1]{$n_{e,\text{sep}}=#1\cdot10^{19}$\,m$^{-3}$}
\newcommand{\MW}[1]{#1\,MW/m$^2$}
\newcommand{\etaln}{ et al.}
\newcommand{\code}[1]{\mintinline{python}{#1}}
\newcommand{\lined}[1]{$#1\cdot 10^{19}$\,m$^{-2}$}
\newcommand{\Dis}[1]{${D}$ = #1\,m$^\text{2}$/s}

\newcommand{\phiis}[1]{$\phi\approx #1$\textdegree}
\newcommand{\phiisless}[1]{$\phi\lessapprox #1$\textdegree}

\graphicspath{{images/}{}}

\providecommand{\bibAnnoteFile}[1]{}
\providecommand{\bibAnnote}[2]{}

\newcommand{\TITLE}{Parametrisation of target heat flux distribution and study of transport parameters for boundary modelling in W7-X}
\newcommand{\TITLEL}{\TITLE} 

\providecommand{\AUTHOR}{David Bold}
\providecommand{\AUTHORL}{\AUTHOR}

\newcommand{\baffle}{target}

\usepackage[pdftex,%
	    pdftitle={\TITLEL},%
	    pdfauthor={\AUTHORL},%
	    colorlinks,%
	    citecolor=black,%
	    filecolor=black,%
	    urlcolor=black,%
	    linkcolor=black]{hyperref}
\begin{document}

\title{\TITLE}

\author{  David~Bold${}^{a}$, Felix~Reimold${}^{a}$, Holger~Niemann${}^{a}$, Yu~Gao${}^{a}$,
  Marcin~Jakubowski${}^{a}$, Carsten~Killer${}^{a}$, Victoria~R.~Winters${}^{a}$ and the W7-X team${}^{a,b}$}

\address{
  ${}^{a}$Max Planck Institute for Plasma Physics, Wendelsteinstr. 1, 17491
  Greifswald, Germany\\
  ${}^{b}$See Klinger et al 2019~\cite{klinger19a} for the W7-X Team
  }
\ead{dave@ipp.mpg.de}
\vspace{10pt}
\begin{indented}
\item[]\today
\end{indented}


\begin{abstract}
Modelling the scrape-off layer of a stellarator is challenging due to the
complex magnetic 3D geometry.  The here presented study analyses simulations
of the scrape-off layer (SOL) of the stellarator Wendelstein 7-X (W7-X) using
the EMC3-EIRENE code for the magnetic standard configuration. Comparing with
experimental observations, the transport model is validated.  Based on the
experimentally observed strike line width, the anomalous transport
coefficients, used as input to the code are determined to around
$0.2\,$m$^2/$s.  This is however in disagreement with upstream measurements,
where such small cross-field transport leads to temperatures higher than
measured experimentally.
\end{abstract}

\section{Introduction}\label{s:intro}
In order to operate fusion power plants based on the magnetic confinement
concept the power flux on the plasma-facing surfaces needs to be controlled to
prevent the overloading of the structures.
Predictive modelling, necessary for design of next-step fusion devices,
requires successful validation via comparison to existing experimental devices to
ensure all important underlying physics is included in the code. 
One of these devices is Wendelstein 7-X (W7-X), an
advanced stellarator with reduced neoclassical
transport~\cite{klinger17a,wolf17a,pedersen17a,klinger19a,beidler21a},
which had its first divertor operational campaign in 2017 - 2018.

\begin{figure}
  \centering
  \raisebox{-0.5\height}{\includegraphics[height=.49\linewidth]{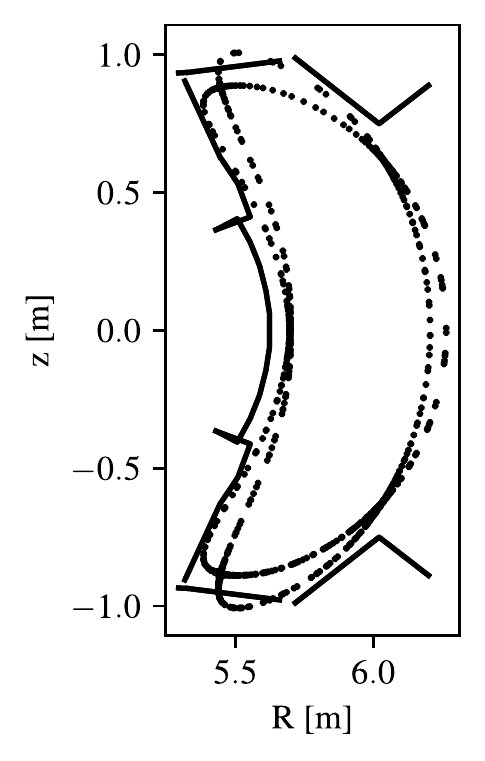}}
  \raisebox{-0.5\height}{\includegraphics[width=.49\linewidth]{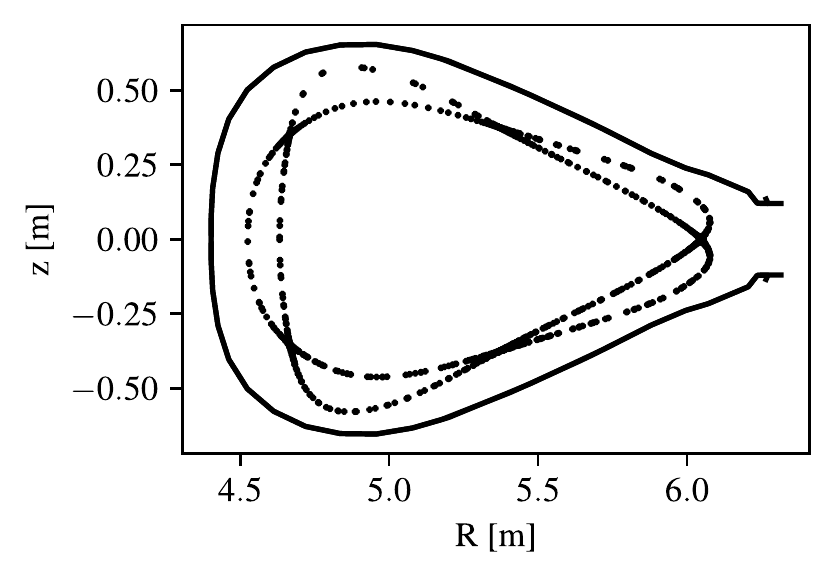}}
  \caption{Shown are the islands at toroidal position $\phi = 0$ (left) and
    $\phi=\pi/5 = 36$\,\textdegree{} (right) as dots as well as the target structures
    used in the simulations.
  }\label{f:poincare}
\end{figure}
In contrast to tokamaks, the scrape-off layer (SOL) of W7-X is inherently
three dimensional.  W7-X features a 5-fold toroidal symmetry. Each of the
5 modules is in itself stellarator symmetric and can be split into two half
modules.
The SOL of W7-X features an island divertor, where in the
standard configuration the 5 resonant islands are intersected by 10
divertor modules~\cite{pedersen18a,pedersen19a,hammond19a}. A plot of the
islands and the intersection with the divertor is shown in
fig.~\ref{f:poincare}.
\todo{@flr: The numbers are used in some plots. I dont think for anything else ...}
The upper halfs of the modules have even
numbers, namely 18, 28, 38, 48, 58, and the lower half-modules have uneven
numbers, 19 to 59. The half-modules $x$8 and $x$9 are in the $x$-th module.
The divertors carry the numbers of the respective half-modules they are
located within.

The lack of toroidal symmetry makes connection and comparison of
experimental measurements at different toroidal locations extremely
complex. As such, there is great demand for 3D modelling, where
synthetic diagnostics can be implemented to help understand whether
differing diagnostic measurements are truly in disagreement or if
differences are due purely to spatial variations in the
plasma~\cite{frerichs16a}.
However, before such an analysis can be performed, it is
critical to first validate the simulations, which itself requires
diagnostic input covering as much of the SOL plasma domain as
possible. 

The anomalous cross-field transport in the SOL of fusion plasmas is
often considered to be dominated by turbulence. In W7-X experiments,
SOL turbulence and turbulent transport has been
observed~\cite{killer20a,killer21b,zoletnik19a,liu19a}.
As fully turbulent simulations of the
full SOL are computationally extremely challenging, simpler models
are generally used, such as fluid transport codes~\cite{winters21a}.
There the turbulence is effectively represented by anomalous diffusion
coefficients.

This work validates the diffusion-based anomalous transport of
EMC3-EIRENE in the absence of drifts in the
model~\cite{feng14a,feng21a} by using spatially
constant transport coefficients. The simulation data is compared with experimental data
from W7-X from the infra-red heat flux diagnostic and the reciprocating electric probes.
This work extends previous work and 
especially addresses remaining discrepancies between simulations and
experiments~\cite{feng21a,feng14a,schmitz20a,wurden17a,schmid20a,effenberg19a,lore19a,winters21a}.
The here presented analysis is restricted to the magnetic
standard configuration.

The current paper is organized as follows: Section~\ref{s:method} the
newly developed methods for comparing the heat-fluxes from experiments
and simulations is presented.
In section~\ref{s:exp} the experimental data is presented, where the
toroidal power distribution and the strike line width is seen.
In the following section, the simulations are presented.
Section~\ref{s:discussion} summarises and discusses the results, here
it is seen that a spatially constant diffusion coefficient cannot
simultaneously match downstream and upstream conditions in the
selected magnetic configuration.
The main conclusions are presented in the final section.


\begin{figure}
  \centering
  \includegraphics[width=.5\linewidth]{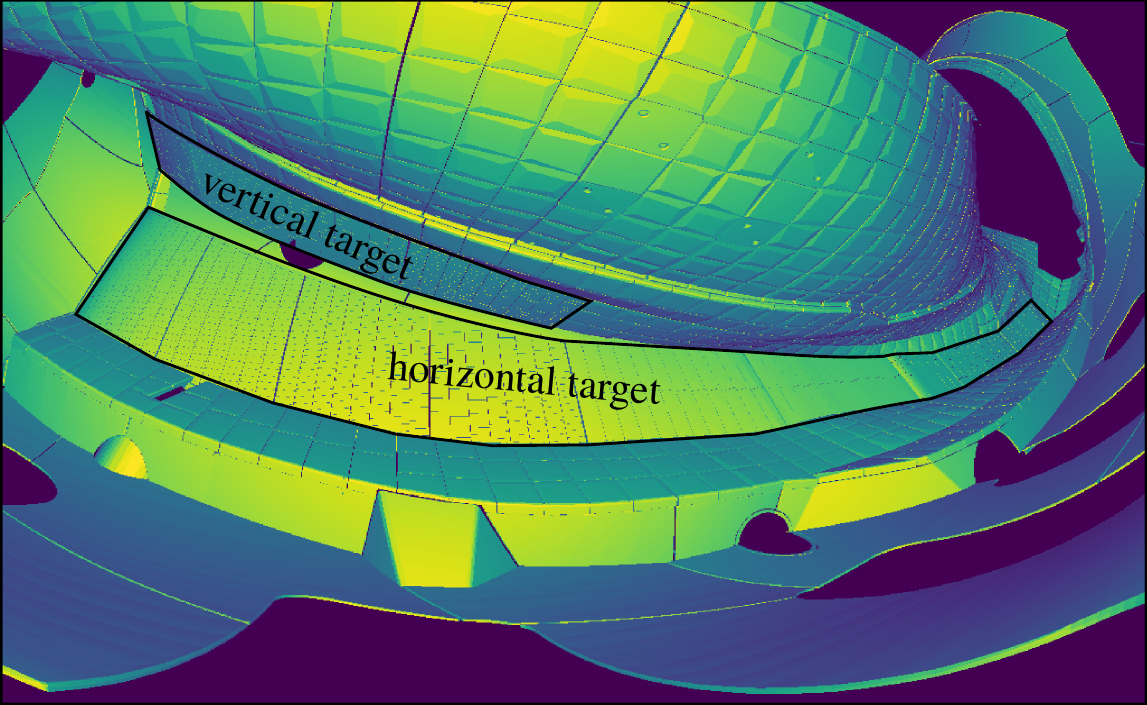}
  \caption{Synthetic view of the IR camera on the target. 
  }\label{f:finger10}
\end{figure}
\begin{figure}
  \centering
  \includegraphics[width=.5\linewidth]{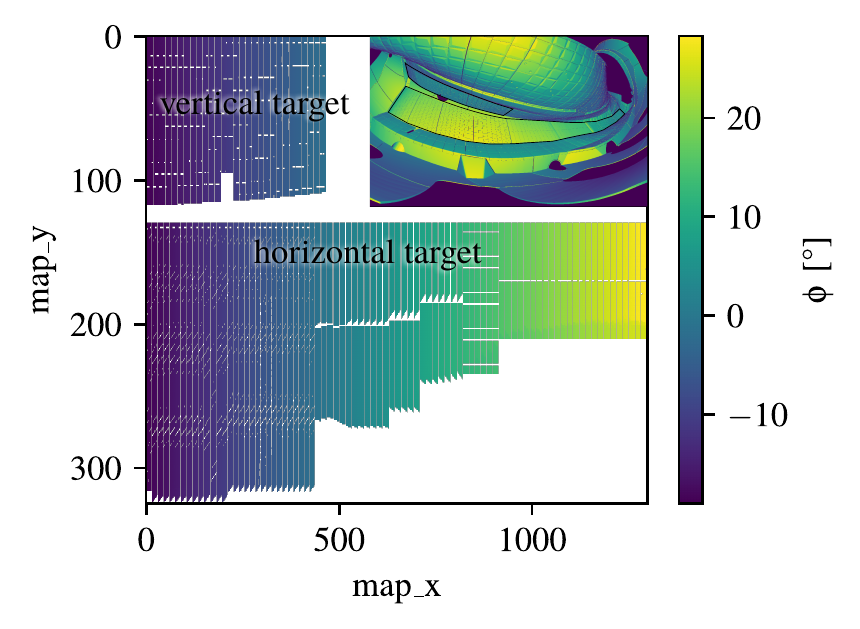}
  \caption{Mapped view of the target. 
    The coordinates \code{map_x} and
    \code{map_y} are not directly related to physical
    quantities. \code{map_x} is roughly aligned with the magnetic
    field, while \code{map_y} is roughly orthogonal to the magnetic
    field. The pumping gap is around $\code{map_y} \approx 125$.
    The finger structures extend roughly 15 pixel in \code{map_x} direction.
    In
    the top right is an inset of fig.~\ref{f:finger10}.
    The so called low iota target is on the horizontal target, for $\phi < 0$.
    Note that all targets are mapped to this $\phi$ coordinate. Upper
    divertors are mapped to negative toroidal angle, due to stellarator
    symmetry. Divertors are also shifted by $\frac{n\cdot 2\pi}{5}$ due to the
    5 fold symmetry of W7-X.
    \todo{additional plot of phi?}
  }\label{f:fingermap}
\end{figure}

\section{Method}\label{s:method}
\todo{8) p.3, Methods: The reader wonders why divertor Langmuir probe data (as
  presented in [7]) are not included in this work at this point. They provide
  important information about downstream/divertor target electron density,
  temperature and heat-flux and may further constrain the modelling - how do
  those profiles compare to the IR data?}
\todo{Include LP or discuss?}
\todo{Add LP to future work}

\subsection{W7-X diagnostics}
In this work, two diagnostics are used for comparison to simulation
data: one is measuring downstream at the divertor targets and the other
upstream. Both downstream and upstream parameter comparisons are
important to determine if the EMC3-EIRENE simulations successfully
reproduce features across the entire SOL.  The downstream measurement
used is the infra-red (IR) camera system~\cite{jakubowski18a}, which
fully covered the area of the 10 divertors in the previous
experimental campaign, with data available for 9 of the 10.

The temperature is derived from the IR
radiation. The heat-flux is calculated by the evolution of the temperature
profiles using the two-dimensional thermal model THEODOR~\cite{sieglin15a}.
This heat-flux is used to assess the validity of the employed SOL
transport model in EMC3-EIRENE as the heat-flux has a high spatial
resolution. The spatial resolution is around 3\,mm the noise level is
around \MW{0.25}.

The second diagnostic used for comparison is the  multi-purpose manipulator (MPM).
The MPM can be equipped with different probe heads to provide profiles
of various plasma parameters such as $T_e$, $n_e$, poloidal mach
number and more~\cite{killer19a,killer21a}.

\begin{figure}
  \centering
  \includegraphics[width=.7\linewidth]{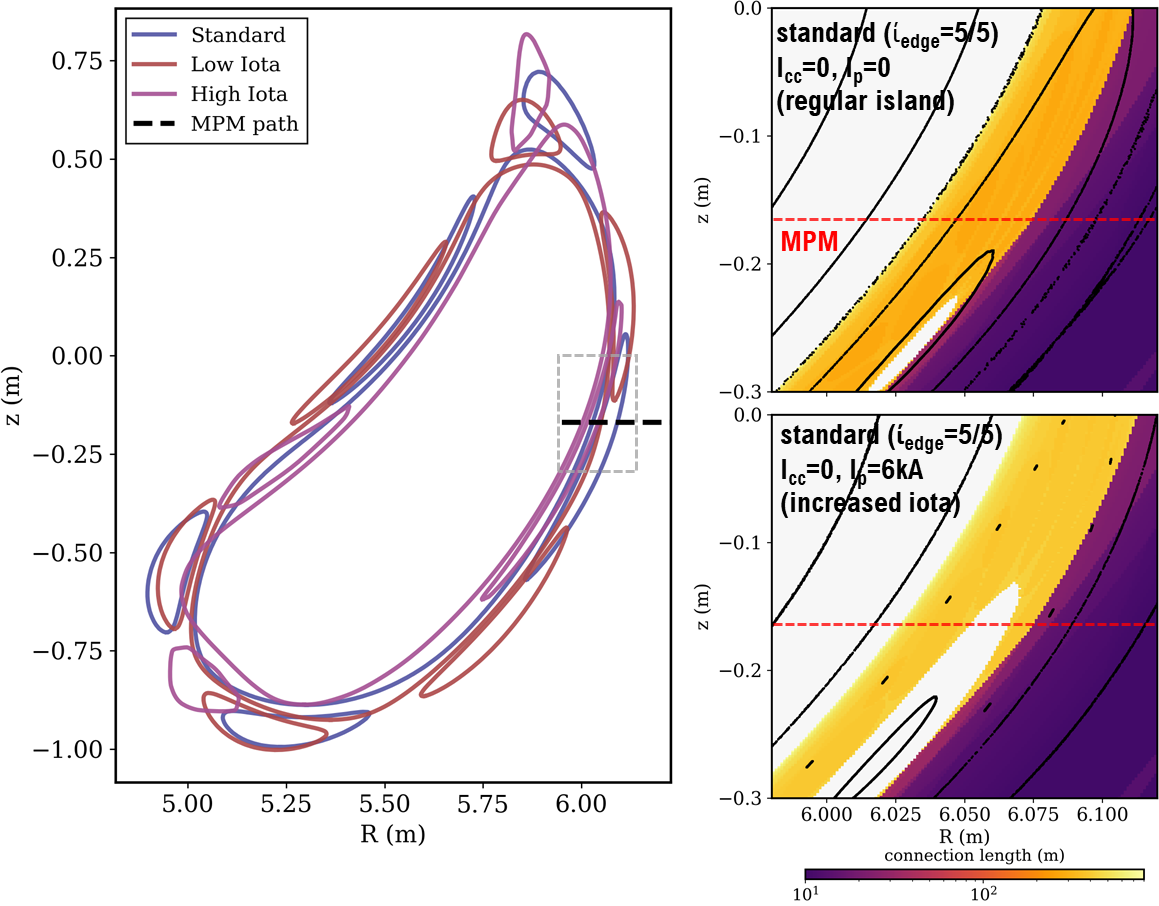}
  \caption{Position of the MPM diagnostic. Reproduced with kind permission
    from~\cite{killer21a}.
    \todo{use nicer plot with only applicable islands}
    \todo{Do not hide island with colorbar}
  }\label{f:mpm:pos}
\end{figure}
In the here presented analysis, the MPM was equipped with Langmuir
probes, that have been used to measure the electron density and
temperature in the SOL of W7-X. This provides plasma parameters
upstream and thus complements the downstream comparison provided by
the heat-flux measurements at the divertor. Unlike the infra-red
diagnostic the MPM is only present in one location, thus does not give
a direct measurement of up-down asymmetries~\cite{hammond19a} or field
errors~\cite{lazerson18a}.
The path of the MPM is shown in fig.~\ref{f:mpm:pos}.

\subsection{Heat-Flux distribution analysis}\label{s:fluxana}
The strike-line width and amplitude is used in order to make the
heat-flux profiles more comparable between modelling and experiments.
The IR data is mapped onto the image format as shown in
fig.~\ref{f:fingermap}.
The divertor is split into smaller structures, called fingers, that
extend mostly in poloidal direction.  1D slices of the data are
analysed, taking slices roughly perpendicular to the magnetic field
lines, in the \code{map_y} direction from fig.~\ref{f:fingermap}.
This gives around 15 1D slices for each finger.
These 1D slices are then fitted to a function consisting of a constant
background plus 2 Gaussian.  The positions of the peaks are
constrained to be within the data slice. The lower bounds for the peak
is 3 times the grid spacing, to avoid fitting a single outlier, rather
than the general shape of the data. For fitting the Trust Region
Reflective algorithm (\code{trf}) from
\code{scipy.optimize.least_squares} is used~\cite{byrd86a,branch99a}.

In order to decrease the computational cost as well as to decrease the impact
of noise, 10 time frames are averaged for fitting. This gives a time
resolution of 100\,ms. The study is restricted to steady-state
profiles as the EMC3-EIRENE code only provides steady-state
solutions.
Although the evolution of the toroidal plasma current throughout the
discharge may change the location and width of the strike line, the
movement is on the order of mm/s and does not significantly impact the
result over a 100\,ms time window~\cite{gao19a}.

For each averaged time slice, each pixel row of each finger is separately analysed.
One should note, that for one \code{map_x} value there can be several data
slices, as the horizontal target and the vertical \baffle{} are different
fingers but share the same \code{map_x} value. Similarly, in the middle
of the horizontal target, between $450 \lessapprox$ \code{map_x}
$\lessapprox 800$ the data is split into two
fingers, one close to the gap, and one further away.  If the peak heat-flux is
below the noise-level of \MW{0.25}\todo{@Yu is there a reference for
  this?}, no fit is attempted.

\begin{figure}
  \centering
  \includegraphics[width=.5\linewidth]{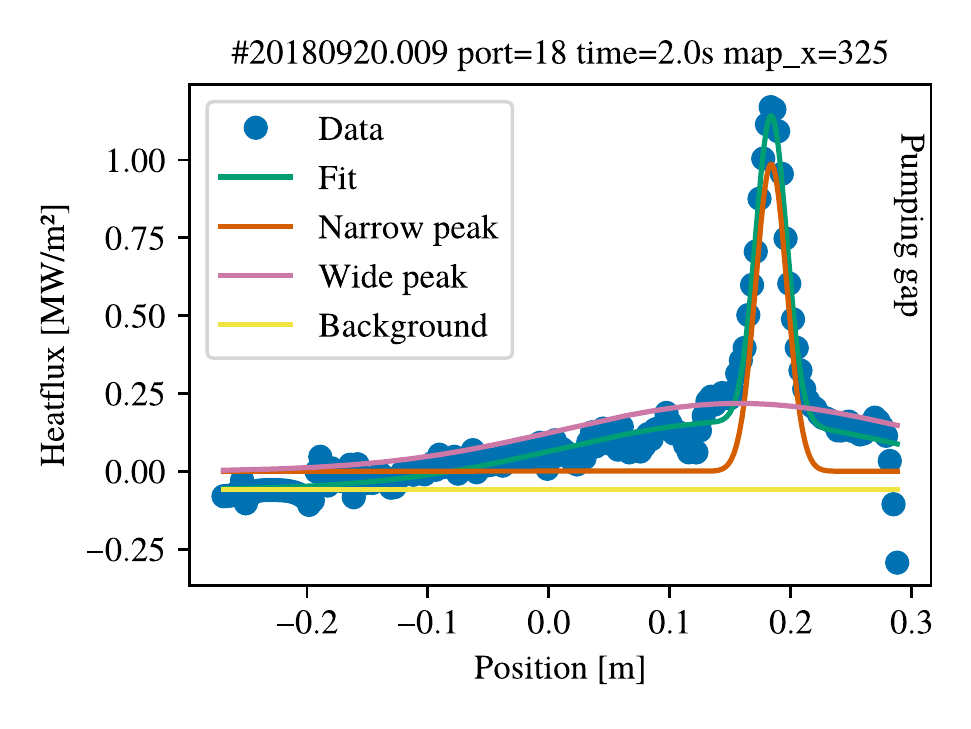}
  \caption{Plot of the IR data for the horizontal target at
    \code{map_x=325}. It can be seen that the data consists of a narrow, high
    in amplitude peak, as well as a broader feature with a significantly
    smaller amplitude. The fitted constant background is in this case
    negligible.
  }\label{f:fit0}
\end{figure}
\begin{figure}
  \centering
  \includegraphics[width=.5\linewidth]{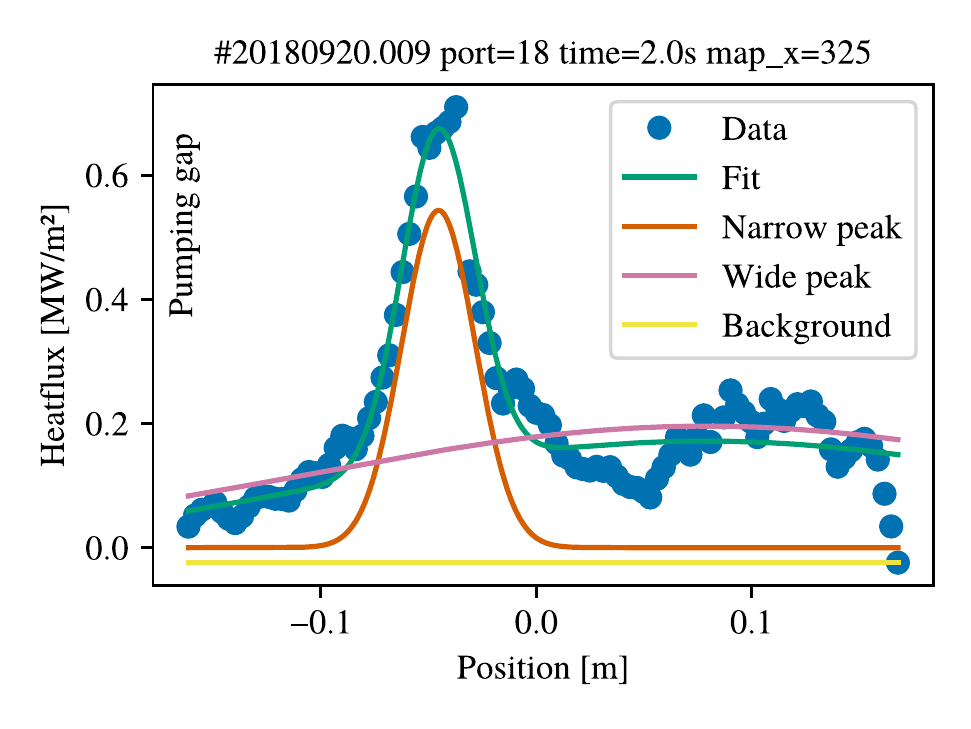} 
  \caption{Plot of the IR data for the vertical \baffle{} at
    \code{map_x=325}. The fit detects the strike line as ``Narrow peak''. The
    ``Wide peak'' combined with the ``Background'' fit the background as well
    the reflections at 0.1\,m.
  }\label{f:fit1}
\end{figure}

Examples of the fitted data are shown in fig~\ref{f:fit0} and
fig.~\ref{f:fit1} for the horizontal target and vertical \baffle{}. Especially
on the vertical target for \code{map_x} $lessapprox 450$ the fit is generally
good. On the middle of the horizontal \baffle{},
especially for the fingers distant from the pumping gap, the heat-flux is very
low, and heat-fluxes above the \MW{0.25} limit are typically a single spike
due to noise, giving a strike-line width of the lower bound $\lessapprox
1$\,cm. A problematic fit is shown in fig.~\ref{f:fit:bad} that will
be discussed later.

On the vertical \baffle{}, shown in fig.~\ref{f:fit1} the second structure at
position 0.1\,m is due to reflections\todo{@Yu: Is there a reference?}, but as
the the analysis is mostly concerned about the more narrow, higher peak, the
reflection does not affect this analysis.

The fitted quantities were averaged by weighting with the power $P_i$ of the data
slice. Only slice $i$ where fitting is attempted are included.
The power-averaged quantity $\alpha$ is thus
\begin{align}
  \alpha = \sum_i \alpha_i P_i / \sum_i P_i
  \label{eq:mean}
\end{align}

\begin{figure}
  \centering
  \includegraphics[width=.49\linewidth]{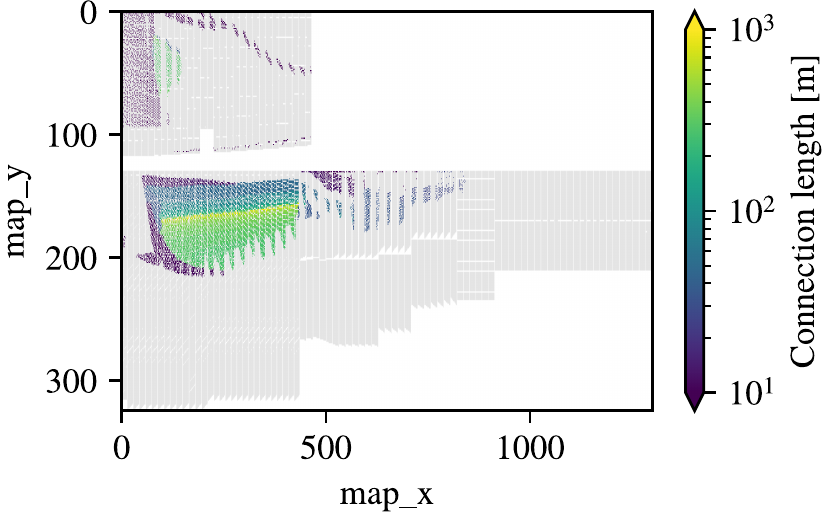}
  \includegraphics[width=.49\linewidth]{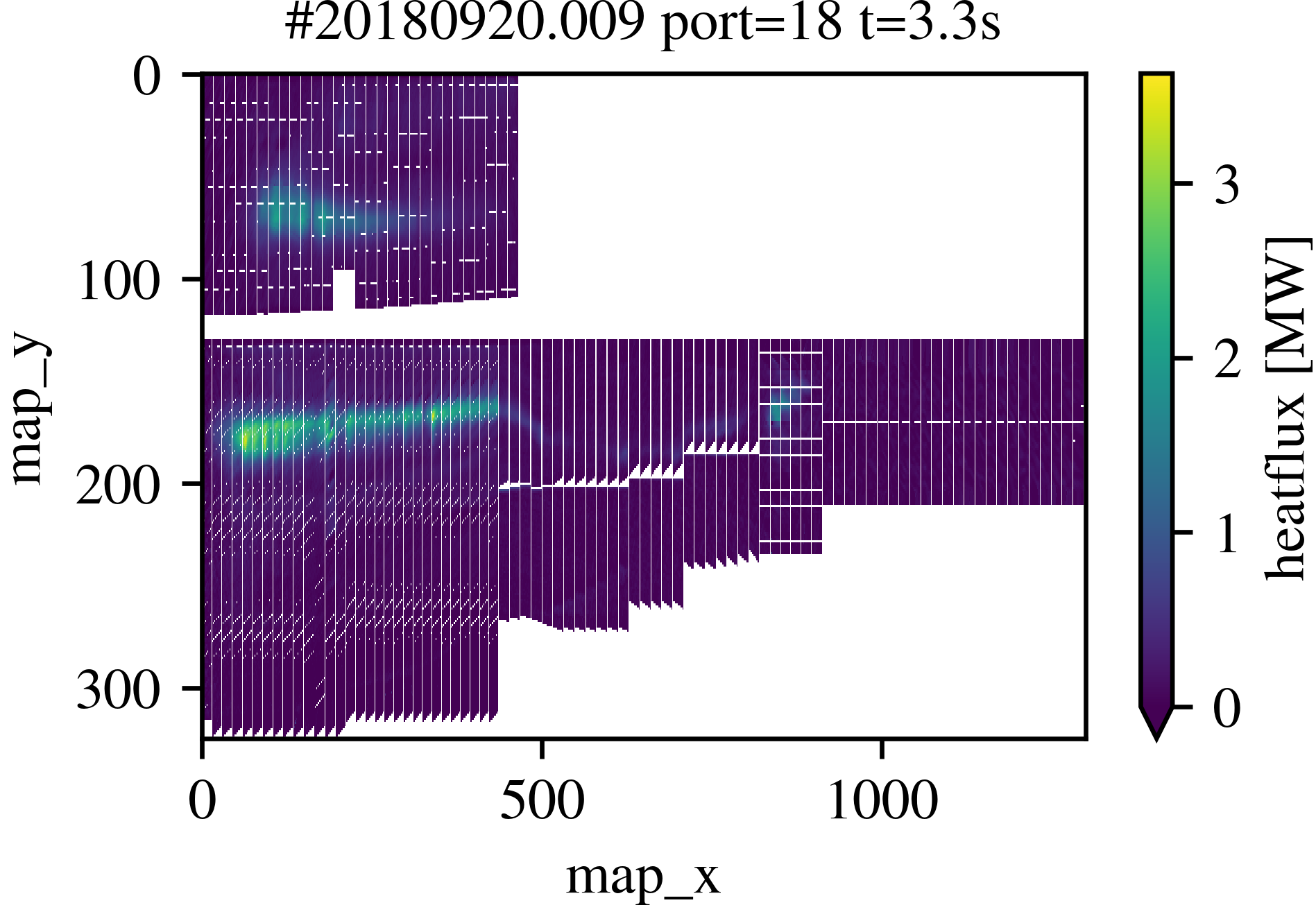}
  \caption{Shown on the left is a plot of the connection length mapped
    on to the target.  Plotted in grey 
    is the target regions where no traced field line ended.
    On the right is a  plot of the heat-flux on the divertor at
    $t=t_1+3.3$\,s for shot \#20180920.009.
    The main strike line is on the left of the horizontal target, roughly in
    agreement with the long connection lengths. Additional
    heat loads on the high iota target as well as the vertical \baffle{} are
    visible.
  }\label{f:con}
\end{figure}

Fig.~\ref{f:con} shows on the left the connection length of some regions of the
target regions.  Regions of very long connection length $> 1000\,$m
indicate the location of the main strike line formed by the
intersection of the island on the divertor target plates.  It can be
seen the main strike line is on the low iota target.  Additionally,
also on the vertical \baffle{} long connection lengths are observed.

\subsection{EMC3-EIRENE}\label{s:emc3}
EMC3-EIRENE is a Monte Carlo fluid transport and kinetic neutral code, that is capable of
handling complex geometries, such as those commonly encountered in the SOL of
stellarators. It has already been used in the past to model the edge of
W7-X~\cite{feng21a,winters21a,feng14a,schmitz20a,wurden17a,schmid20a,effenberg19a,lore19a}.
While EMC3-EIRENE does captures some of the observations in experiments,
especially global trends~\cite{winters21a,feng21a}, there is still
disagreement in local parameters\cite{feng21b}.

EMC3-EIRENE does include parallel transport in the form of advection as well
as viscosity and parallel heat diffusivity\todo{add equation?}.  Perpendicular transport included
in EMC3-EIRENE features anomalous diffusion based on some given particle and
heat diffusion coefficients. EMC3-EIRENE does not require nested
flux-surfaces and is only aware of the local magnetic geometry.  For this
reason the perpendicular diffusion is uniform in radial and bi-normal
direction, i.e. $D \propto I - \vec b \vec b$ with $\vec b$ the unit vector in
the direction of the magnetic field and $I$ the identity matrix. Drifts, like
the $E\times B$ drift, are not included in EMC3-EIRENE. While EMC3-EIRENE is
able to handle spatially varying perpendicular transport
coefficients~\cite{feng14a,zhang16a}, this feature is not used in the here
presented study.

An analysis analogous to the one described in sec.~\ref{s:fluxana} can be applied
to simulated heat-flux data generated by EMC3-EIRENE. Thus, a direct
comparison of experiment and modelling is performed.
This allows to quantify the discrepancies and validate the modelling
and the assumptions, for example the transport model, with
experimental data.

In all cases the heat diffusion coefficient $\chi$ is set to $\chi =
3\cdot D$, i.e. scaled with the diffusion coefficient. Note that the
resulting heat transport is $q_\perp \propto n \chi$, i.e. has a density
dependence even for constant $\chi$.

\subsection{xemc3}
The majority of the analysis has been carried out using the xarray
framework~\cite{hoyer17a,xarray_0_17_0}.
For that the xemc3~\cite{xemc3-0.1.0} library has been implemented that reads
the output of the EMC3-EIRENE routine into the xarray format. An extensive
documentation, including documentation and online tutorials, is available
online~\footnote{\url{https://xemc3.readthedocs.io/}}.

\section{Experimental data}\label{s:exp}
\begin{figure}
  \centering
  \includegraphics[width=.5\linewidth]{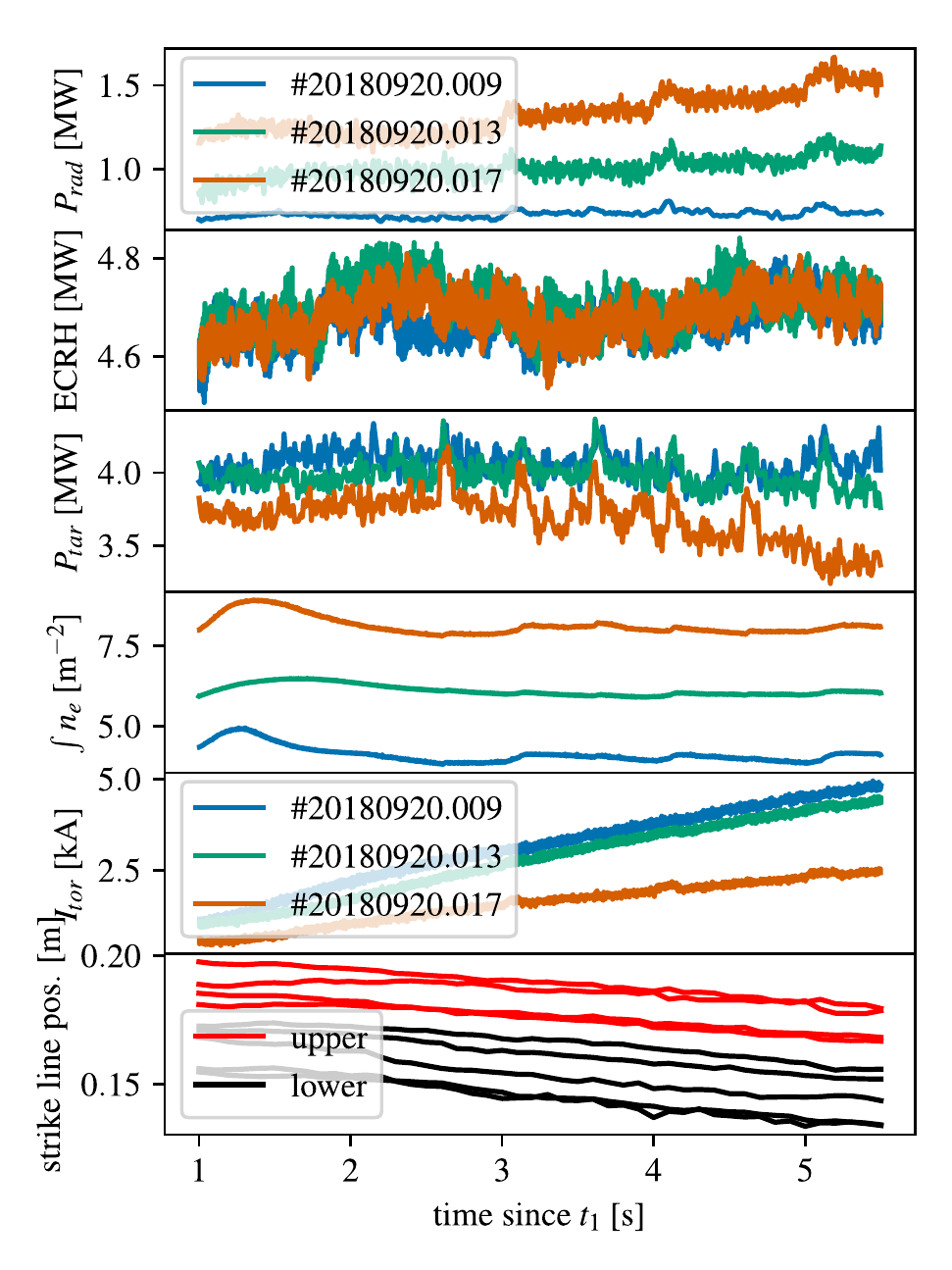}
  \caption{Overview of the time evolution of the radiated power $P_{rad}$, the
    ECRH heating power, the power on the target measured by IR, the line
    integrated density $\int n_e$, the toroidal plasma current $I_{tor}$ and the
    strike line position from \#20180920.009 on finger 24. For the target
    modules where no IR data was available, the average from the other targets
    was used to extrapolate to the total target power.
  }\label{f:exp:over}
\end{figure}

For this analysis the W7-X experiments \#20180920.009, 
\#20180920.013 and \#20180920.017
have been analysed. They are part of a density scan with an input power of
4.7\,MW ECRH. They have been selected due to the low radiation
fraction $f_{rad}$ of around $0.15 \ldots 0.35$.  Low $f_{rad}$ avoids
large effects of power dissipation in the volume.  Thus transport is
prominent and easier to study.  The heat-flux on the divertor measured by IR was $3\ldots 4$ MW, shown
in fig.~\ref{f:exp:over}, with the time-averaged power per target between 330\,kW
and 496\,kW.  The peaks in the time evolution, shown in fig.~\ref{f:exp:over}, are due to CH${}_4$ puffs and fuelling.  The magnetic configuration used was the standard
configuration. The toroidal plasma current increased over time and reached
$\approx 5$\,kA after around 6 seconds towards the end of the discharge, with the
exception of \#20180920.017, where the maximal bootstrap current was around
2.5\,kA.
The SOL of W7-X is sensitive to plasma currents and the toroidal plasma
current impacts the heat deposition~\cite{geiger10a,geiger10b,lore19a}. For
the here analysed discharges, the strike-line
width is not significantly impacted by the toroidal current, but the
strike-line position is a function of plasma current~\cite{gao19a}, as shown in
fig.~\ref{f:exp:over}.

The line integrated density was \lined{4} to 
\lined{8}. These low to medium density cases were selected as they feature a
low radiative fraction. 
This allows to focus on the heat transport effect on
the target heat load distribution,
reducing the additional impact of radiation, simplifying the required physics to model
the dynamics and reducing the system complexity as $P_{rad}$ is a strong
function of the electron temperature $T_e$. As the heat-flux is proportional
to the density $q_\perp \propto n \chi$, a density scan was chosen for this study.
The simulations do not feature the same densities, as in the simulations the
separatrix density was set, while in the experiment the separatrix
density is not well known. On the other hand the line integrated
density is not known in the simulations, as the core region is not modelled.

Fig.~\ref{f:con} (right) shows an example of the spatial distribution of the
target heat-flux in the projection described in fig.~\ref{f:fingermap}.
Only the strike line on the low iota target is expected from simple field line
tracing. The load on the vertical \baffle{} can be explained by
field line tracing in the reverse direction\todo{@flr, do you have a
  reference? I can see this from my field line tracing ...},
while other loads are only possible due to cross-field transport.

\subsection{Toroidal distribution}
\begin{figure}
  \centering
  \includegraphics[width=.9\linewidth]{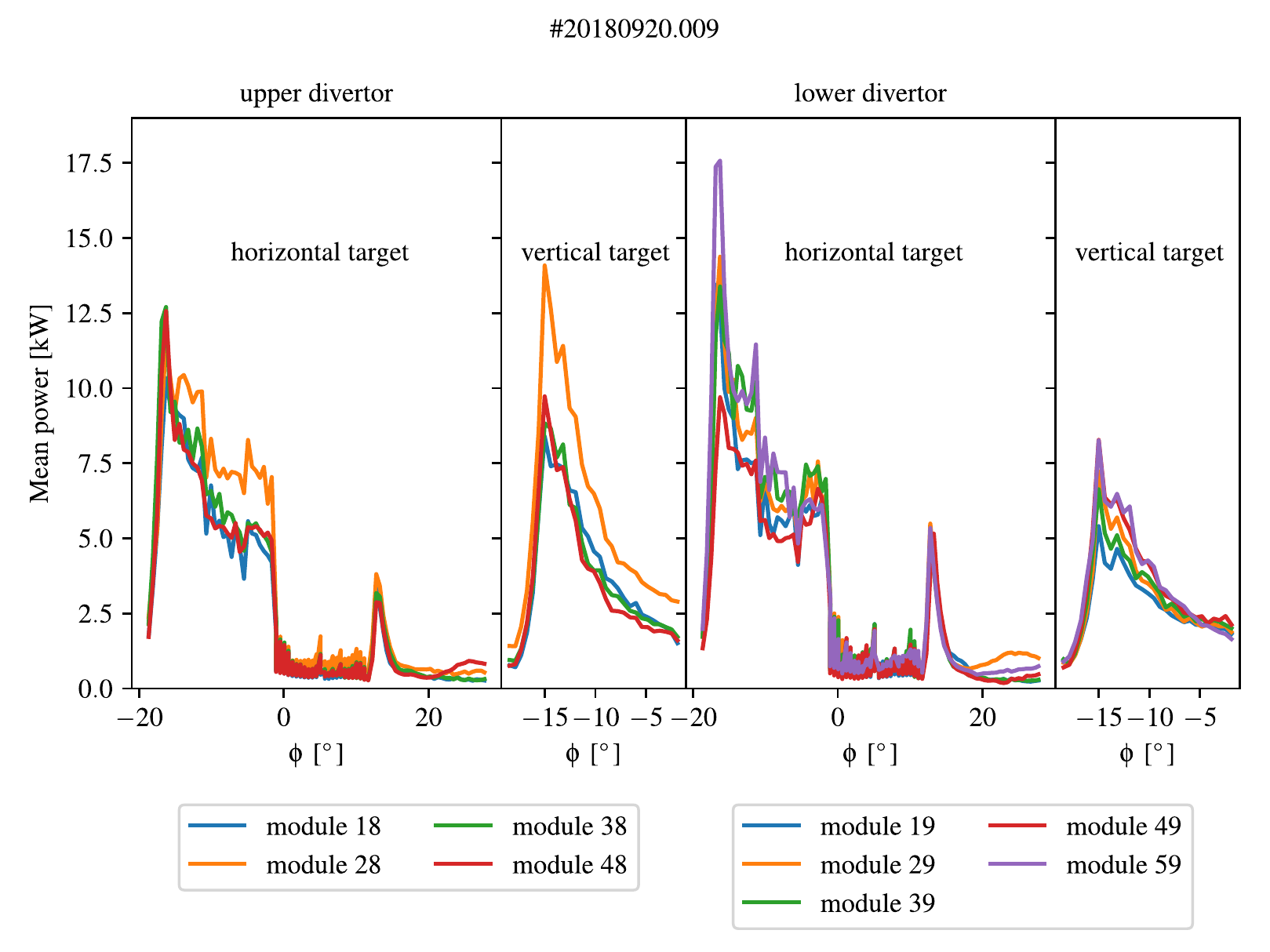}
  \caption{Plot of the time averaged power per finger for the
    steady-state phase of \#20180920.009.  The error bars denote the
    standard deviation of the time evolution. The fingers
    are introduced in fig.~\ref{f:finger10} and
    fig.~\ref{f:fingermap}. Note that the alternating structure in the
    middle of the horizontal \baffle{} is caused by the numbering, even numbers are
    close to the pumping gap, while the odd numbers are further away.
    The different half modules are shown separately. No data from
    half-module 58 is available.
  }\label{f:powerfingersingle}
\end{figure}

Fig.~\ref{f:powerfingersingle} shows a plot of the toroidal
distribution of the heat-flux, by showing the mean power on the
respective fingers, as introduced in fig.~\ref{f:finger10} and
fig.~\ref{f:fingermap}.

No strong variation for the different half modules is observed, only
half-module 28 shows an increased heat-flux at \phiis{-5} on the
horizontal target, as well as on the vertical \baffle{}. For the lower
divertors, most variation is observed at \phiis{-15} where
module 59 shows an increased heat
load and module 49 shows a decreased heat load. These variations might be
explained by field errors~\cite{lazerson18a}.  The calibration of the
absolute values of the IR diagnostic was incomplete in OP
1.2b\todo{@Yu: is there a reference?}.
This limits the reliability
of comparisons between the different IR cameras and thus between the different
half modules.
The simulations assume stellarator
symmetry and therefore only one half-module is modelled.  They are
inherently up-down symmetric and no variation
between different modules is included. For these reasons the following
analysis will focus on averages of the different modules.

\begin{figure}
  \centering
  \includegraphics[width=.9\linewidth]{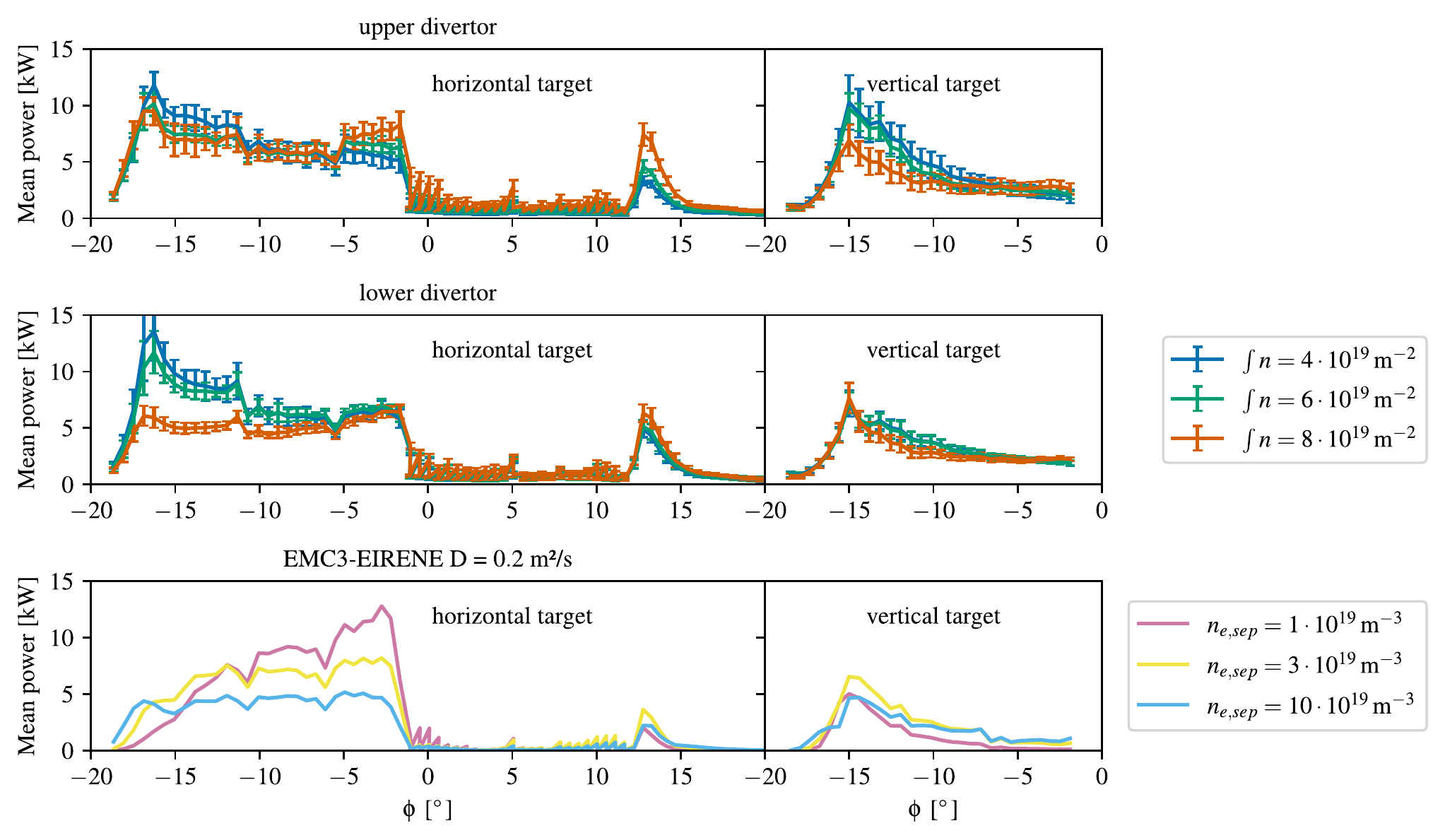}
  \caption{Plot of the time and module averaged power per finger for
    the steady-state phase of \#20180920.009, \#20180920.013 and
    \#20180920.017 on top for the upper divertors and in the middle
    for the lower divertors. On the bottom are results from
    EMC3-EIRENE simulations with \Dis{0.2} discussed in sec.~\ref{s:simulations}. The
    error bars denote the standard deviation of the time evolution and
    inter module variation. 
    As the simulations don't have a
    time component, only the experimental data has error bars.
    The $P_{rad}$ for the simulations was $1\,$MW and for
    the experiments 0.68\,MW to 1.7\,MW, see fig.~\ref{f:exp:over}. The power on
    the divertor was 329\,kW, 306\,kW and 254\,kW for the \dens{1}, \dens{3}
    and \dens{10} case respective. Note that simulations and experiments do not
    match in density, as for the experiments the separatrix density is not
    well known, and for the simulations, due to the lack of core profiles, the
    line integrated density is not known.
  }\label{f:powerfinger}
\end{figure}
Fig.~\ref{f:powerfinger} shows the experimental power per finger that was measured in the
steady state part of a density scan for the upper divertors (top) and
the lower divertors (middle).  In general, a decreasing trend of power
on the target with increasing density is observed, which is expected in the
experiments as with increasing density the radiation increases, and thus the
target heat load is reduced.
An exception to the decreasing trend is
the load at \phiis{12}.
The increased heat flux at this shadowed area is in agreement with an
increased cross field transport with increasing density.
Consequently the load on the middle of the horizontal \baffle{} is increased, at least on the
upper divertors.  At low densities the main heat load on the horizontal target is
mainly at \phiis{-15} and less pronounced at \phiis{-5}.
With increasing
density the ratio of power at \phiis{-15} over the power at
\phiis{-5} is reduced,
suggesting an increased transport channel or a decrease in the losses
from the transport channel to \phiis{ -15}.
Especially on the upper divertors, this results in an increased
heat-flux at \phiis{-5}. At the same time the
heat-flux at \phiis{-15} decreases. Note that, as can be seen on
fig.~\ref{f:con}, the target \phiis{-15} (\code{map_x}$<100$) is shadowed
while at \phiis{-5} (\code{map_x}$\approx 400$) is
directly connected. As such this is in contrast to the expected
behaviour.

Simulations with \Dis{0.2}, shown in the bottom plot
in fig.~\ref{f:powerfinger}, do not see the same trends with density.
Here the peak of
the mean power appears at \phiis{-5} at low
density, while \phiis{-15} generally sees lower
power. As the separatrix density is increased, the mean power at the \phiis{-5} decreases, while the power at \phiis{-15} increases slightly.

\subsection{Strike-line}
For the data shown in fig.~\ref{f:powerfinger} the strike-line has been
analysed using the method discussed in sec.~\ref{s:fluxana}.  Each module and
time-slice has been analysed separately, to not broaden the strike line by
averaging strike-lines at different positions due to the strike-line
movement during the plasma discharge and e.g. camera misalignment and
field errors~\cite{lazerson18a} for different half modules.

The narrow feature identified is expected to be due to parallel plasma
flow to the target. It is not yet clear what is causing the broad
feature.

\begin{figure}
  \centering
  \includegraphics[width=.48\linewidth]{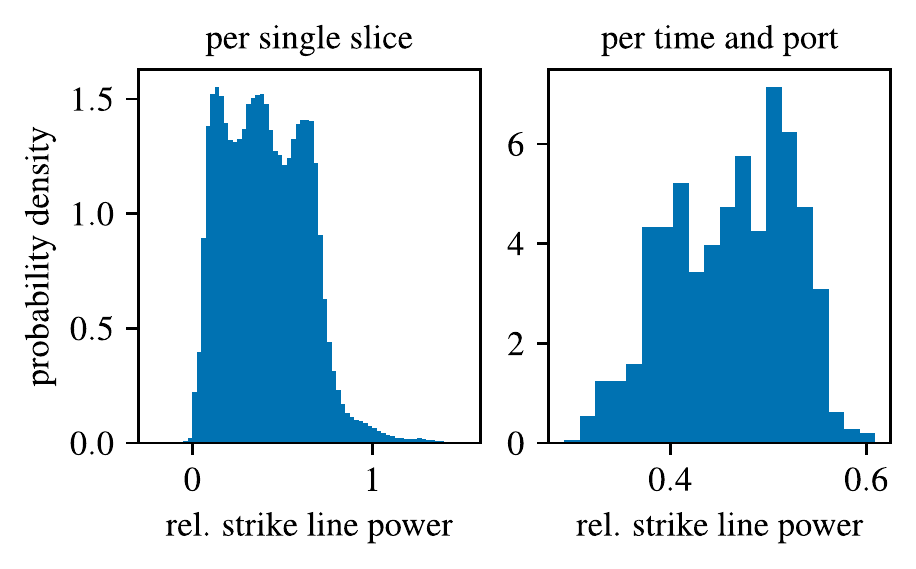}
  \caption{Distribution of the power in the main strike line compared to total power. On the left
    is the histogram for each single fitted slice, while on the right is the
    average for each divertor at each time. The shown data is for experiments
    \#20180920.009 and \#20180920.013.
  }\label{f:power:sl}
\end{figure}
By integrating over the Gauss of the narrow feature, the power of the main
strike line can be calculated. Fig.~\ref{f:power:sl} shows the power observed
in the narrow feature compared to the total observed power.  Roughly 50\,\% of
the power on the divertor is in the main strike line. For the left figure,
``per single slice'' - values below 0 and above 1 are observed. This is due to
bad fits, which can be caused by single points of high heat flux, that are
fitted by a broader Gaussian. However, they are not frequent, and as such it
is expected that they do not have a significant impact. The distribution for
the averaged power on the right is not showing this behaviour, verifying that
this indeed only outliers that show this behaviour.

\begin{figure}
  \centering
  \includegraphics[width=.49\linewidth]{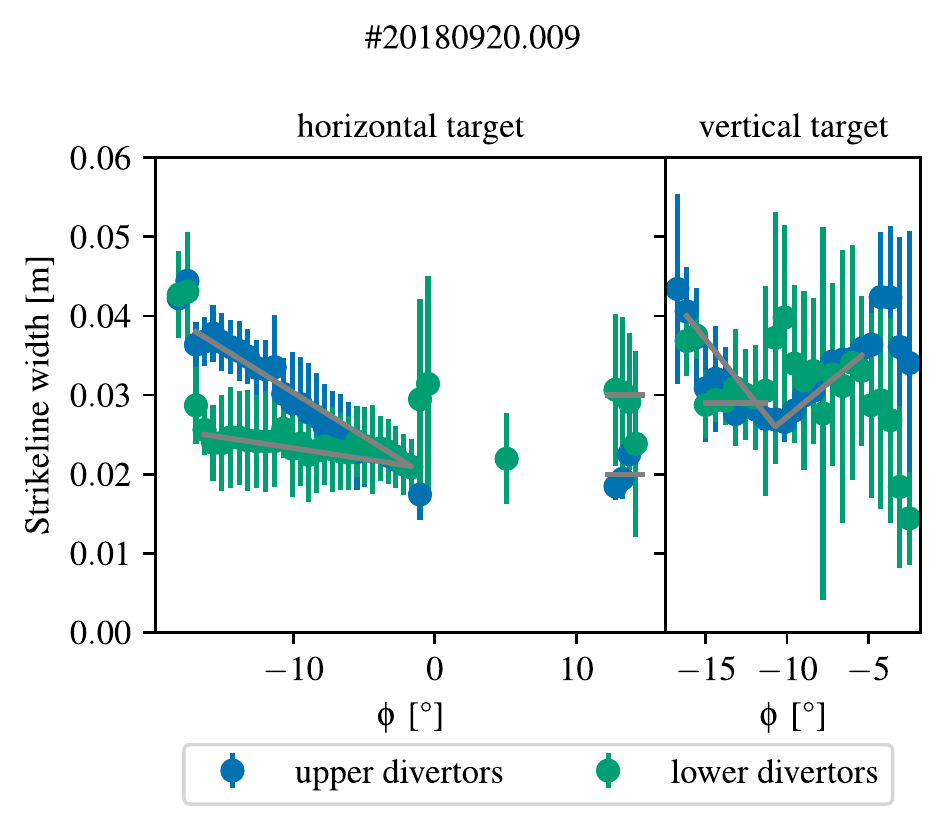}
  \includegraphics[width=.49\linewidth]{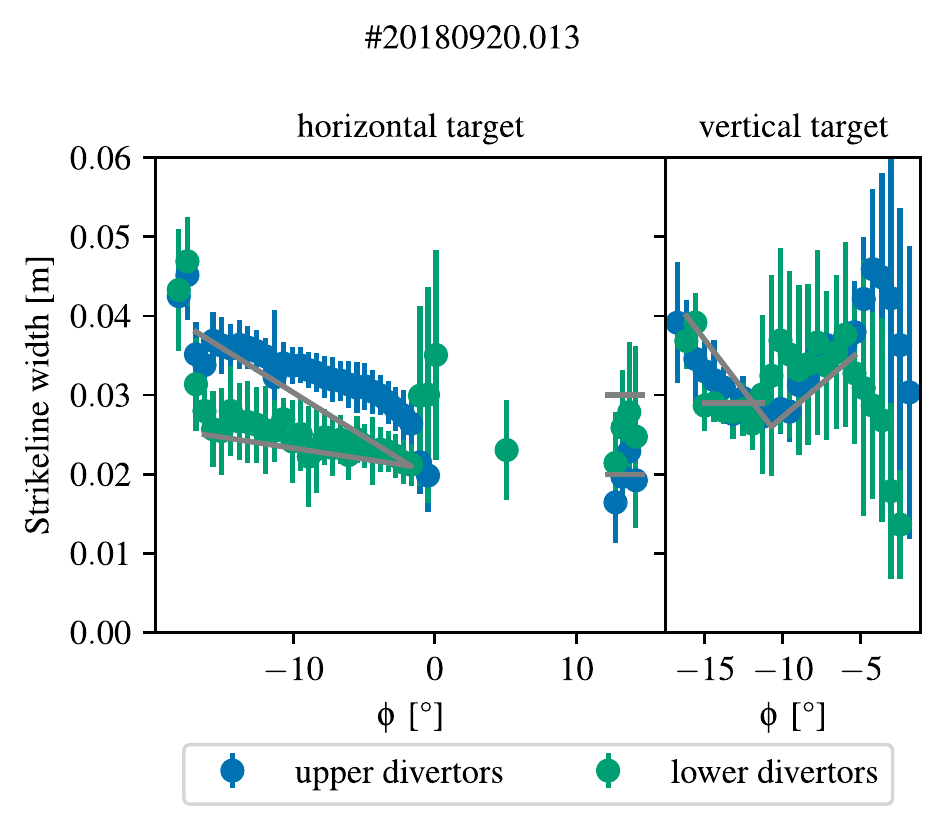}
  \caption{Plot of the time and module strike-line width per finger for
    the steady-state phase of \#20180920.009 (left) and \#20180920.013
    (right).
    The points denote the power-averaged
    strike-line width for all fits, and the error-bars denote the
    power-averaged standard deviation.
    Shown is the narrow strike-line, that has been identified by
    fitting.  Only finger with at least 2\,kW average power are
    shown.
    The grey lines show estimates for \#20180920.009 for upper and
    lower divertors, that will later be used for comparison to
    simulations.
    The large variation in the lower target on the vertical
    \baffle{} is due to significant variation in the time traces for
    some of the fingers.
  }\label{f:slw:09}
\end{figure}

Fig.~\ref{f:slw:09} shows the power-average (see \eqref{eq:mean}) of the
width of the fitted narrow peak 
(see fig.~\ref{f:fit0}) for the low and medium density cases \#20180920.009
and \#20180920.013.
It can be seen that
the strike-line width is in the range of 2\,cm to 4\,cm.\todo{compare
  to holger's results?}

Both density cases show a similar behaviour in their strike-line width
pattern on the upper and lower divertor target plates. Starting
from \phiis{-20}, both the upper and lower
divertor targets see comparable strike line widths and heat flux
magnitudes.
For the remainder of \phiisless{0},
on the lower divertor a narrow strike line is measured for the region
of the high heat-flux, that stays constant, where the power flux is reduced, while on the upper
target the strike-line is broader at \phiis{-15}, end gets more narrow
towards the \phiis{-5}.

The strike-line width on the vertical \baffle{} is comparable on the
upper and lower divertors, at least for the points with significant
power flux.  The vertical targets of the upper and lower divertors
mainly differ in their magnitude of the heat flux, as shown in
fig.~\ref{f:powerfinger}.
The strike-line width at \phiis{15} is slightly wider on the lower
divertor, however they are still the same within the standard
deviation.

\begin{figure}
  \centering
  \includegraphics[width=.5\linewidth]{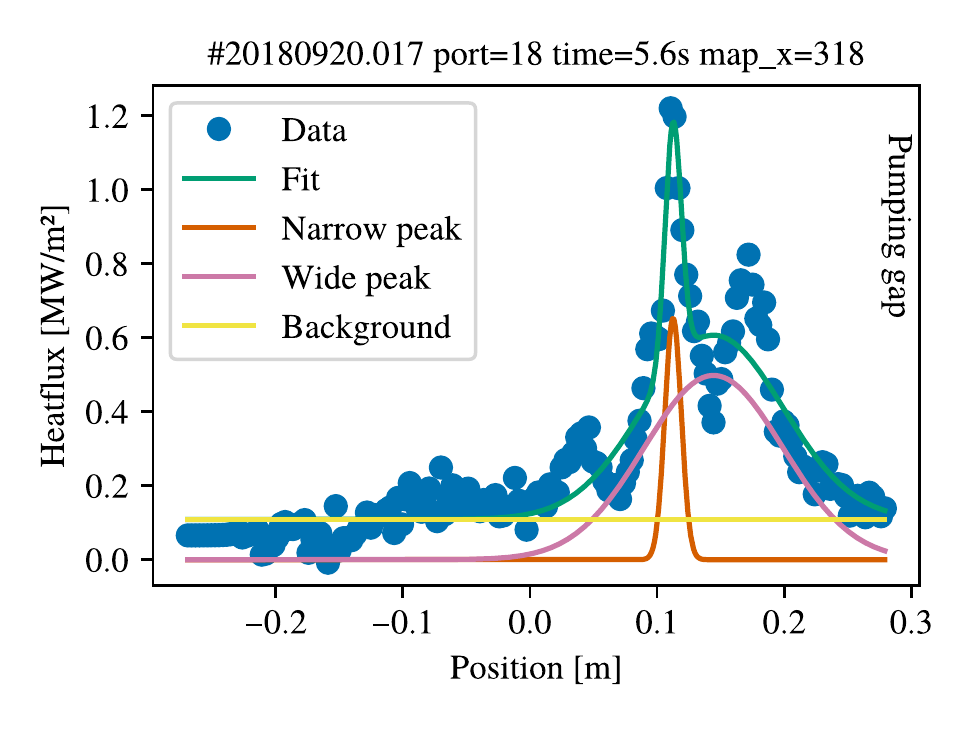}
  \caption{Plot of the IR data for the low iota target and
    \code{map_x=325}. Similar to fig.~\ref{f:fit0} but for like \#20180920.017
    instead of \#20180920.009. Smaller structures in the fit prevent the
    expected convergence of the fit, and the fitted narrow peak is actually a smaller
    perturbation of the true main peak.
  }\label{f:fit:bad}
\end{figure}
For the highest density case \#20180920.017 (not shown), the average strike line
widths across the divertor target could not be computed. Small-scale
structures in the strike-line pattern keep the fits from converging
reliably, with an example shown in fig.~\ref{f:fit:bad}.
The cause of these structures is not yet known. The small scale
structures seem to be fixed in space, while the main strike line moves
in time.  As such it seems
unlikely that the heat-flux of the plasma onto the target does
contain such small-scale structures. It is currently hypothesized that
these structures are caused by artefacts in the IR
diagnostic.\todo{@flr: I haven't looked at this issue more thoroughly
  to see when it happens. It should be fairly easily to download loads
  of data, fit it, and look for bad fits / cases with high
  stddev. Might be something for the ToDo list ...}
For example surface layers could modulate the radiation.
Future work includes understanding the origin of these structures as
well as extending the current fitting mechanism to be
more tolerant of these modulations, such that reliable results of the
strike line features can still be obtained even with their presence.

\begin{figure}
  \centering
  \includegraphics[width=.6\linewidth]{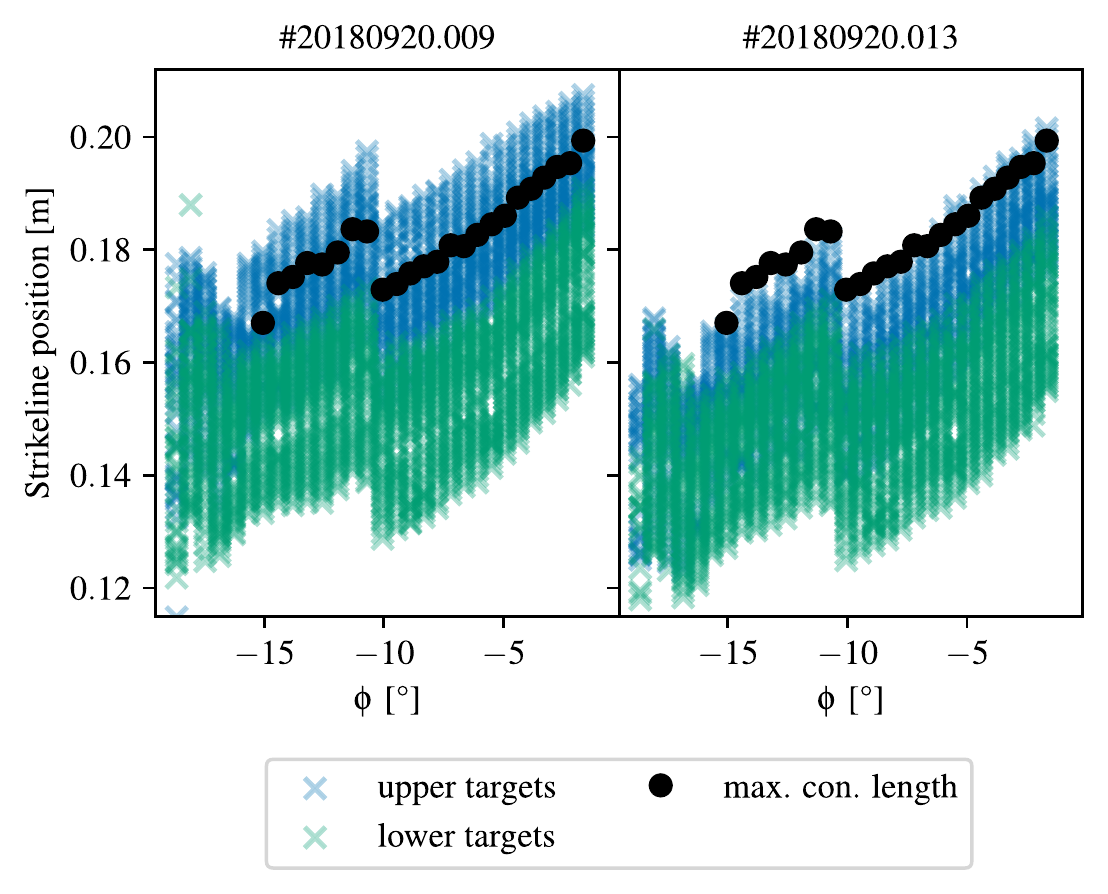}
  \caption{Poloidal position of the strike-line for some slices on the
  horizontal target, time
    averaged between 1 and 2 seconds. The different divertors are colour-coded
    based on the vertical position of the divertor, i.e. upper versus lower
    divertor. Note that the strike line position is not plotted as a function
    of time, but rather as a function of toroidal angle. Thus the jump
    around \phiis{-10} is due to the definition of the local coordinate system.
    Also shown is the position of the long connection length from
    fig.~\ref{f:con} (left). Note that a deviation of the observed strike line
  from the position of the long connection length is expected due to the finite
    toroidal current.
  }\label{f:slp}
\end{figure}
Fig.~\ref{f:slp} shows the poloidal position of the strike-line for a part of
the horizontal target.
The lower divertors tend to have the strike line located somewhat closer to
the pumping gap as compared with the upper divertors.

\subsection{MPM data}
For the above studied discharges no MPM data is available. Thus MPM data from
\#20181010.008, \#20181010.016, \#20181010.021 and \#20181010.022 is used
instead. The discharges used similar input power, same magnetic configuration
and also feature low radiation fraction. 

\section{Simulations}\label{s:simulations}
The scrape-off layer of W7-X has been modelled using
EMC3-EIRENE. For this the upstream density was scanned.
The simulation relies on the stellarator symmetry of W7-X, and
therefore only one half-module is modelled. Ideal coils are used and
thus no error field effects are included. Drifts are not included as
they are not yet implemented in the code.

The input heating power within the simulation domain of one half-module
was set to be 470\,kW, leading to a total of 4.7\,MW for the whole
device.
The power was
distributed evenly between ions and electrons, and enters the domain at the
core boundary. The observed power on the divertor is up to 352\,kW -
giving a total power of $\approx 3.5$\,MW on all divertors.
The upstream density was set to be fixed \dens{1\ldots10}. The cases
\dens{1} and \dens{3} are roughly in the range of the experiments,
while \dens{10} is a purely hypothetical case, as for such high
densities the radiation fraction would be much higher.  No
pumping and fuelling is included in the simulations, and therefore
particle balance is achieved via scaling the recycling flux to the
amount needed for the fixed upstream density value. 
The radiation was fixed to 1\,MW, achieved via carbon impurity radiation, giving a radiation
fraction $\approx 21$\,\%. While in the experiment the radiation fraction
varies from $0.15$ to $0.35$, this was done to study the density and
diffusion coefficient rather
than the influence of the radiation fraction, which was recently studied by
Feng\etaln~\cite{feng21a}. In particular the low $f_{rad}$ was selected to
avoid a dominant effect of the radiation.

The same magnetic field configuration was used in simulation as in
experiment: the standard magnetic field configuration.
A scan in density and diffusion coefficient was performed.
EMC3-EIRENE
stores where and how many particles leave the plasma domain.
These particles can then be mapped onto the target surfaces in a
post-processing step.
In a next step the data has been mapped to the
representation introduced in fig.~\ref{f:fingermap}. From this step on
the same analysis, described in sec.~\ref{s:fluxana}, has been used
for the simulated data as for the experimental data.

\begin{figure}
  \centering
  \raisebox{-0.5\height}{\includegraphics[width=.4\linewidth]{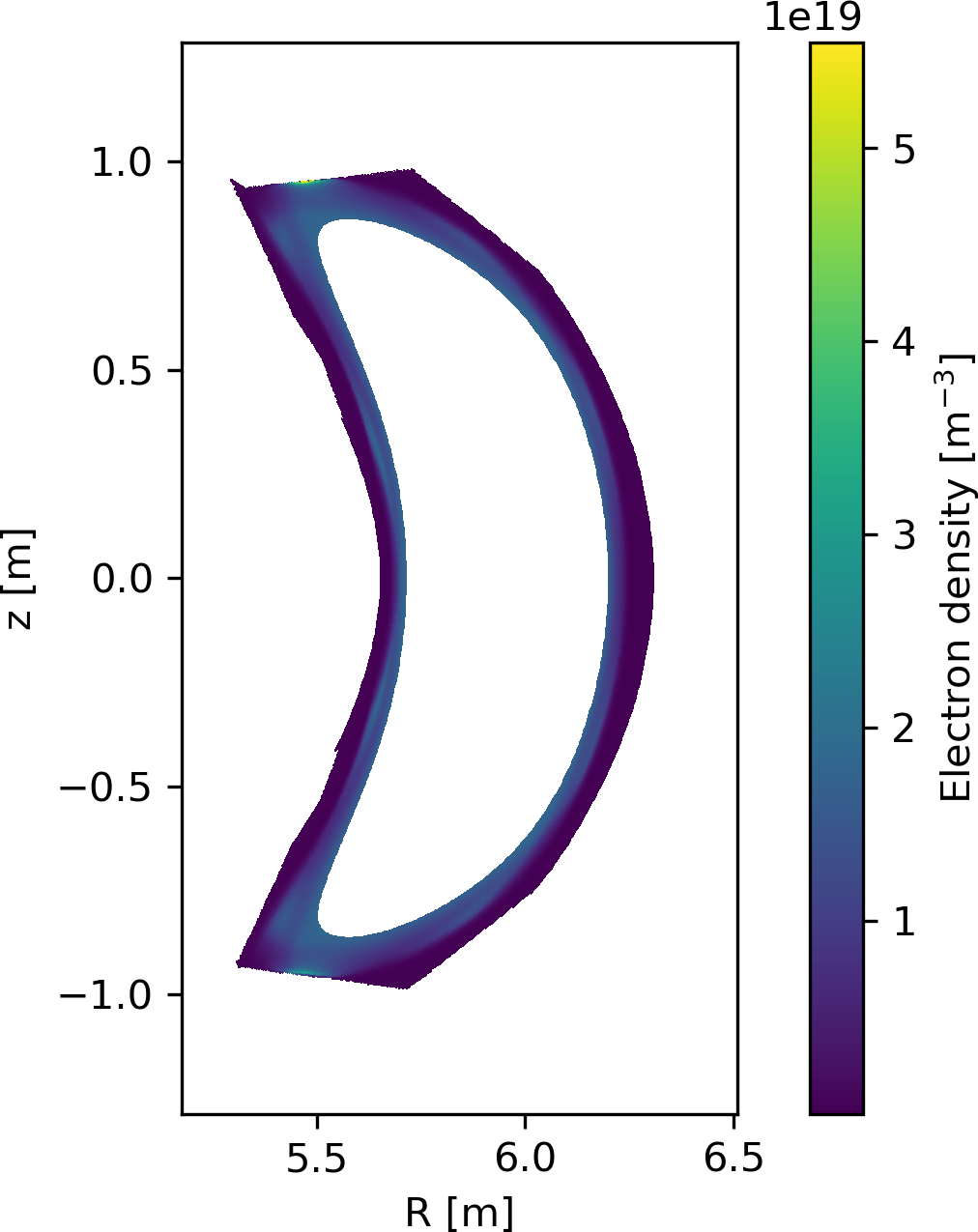}}
  \raisebox{-0.5\height}{\includegraphics[width=.5\linewidth]{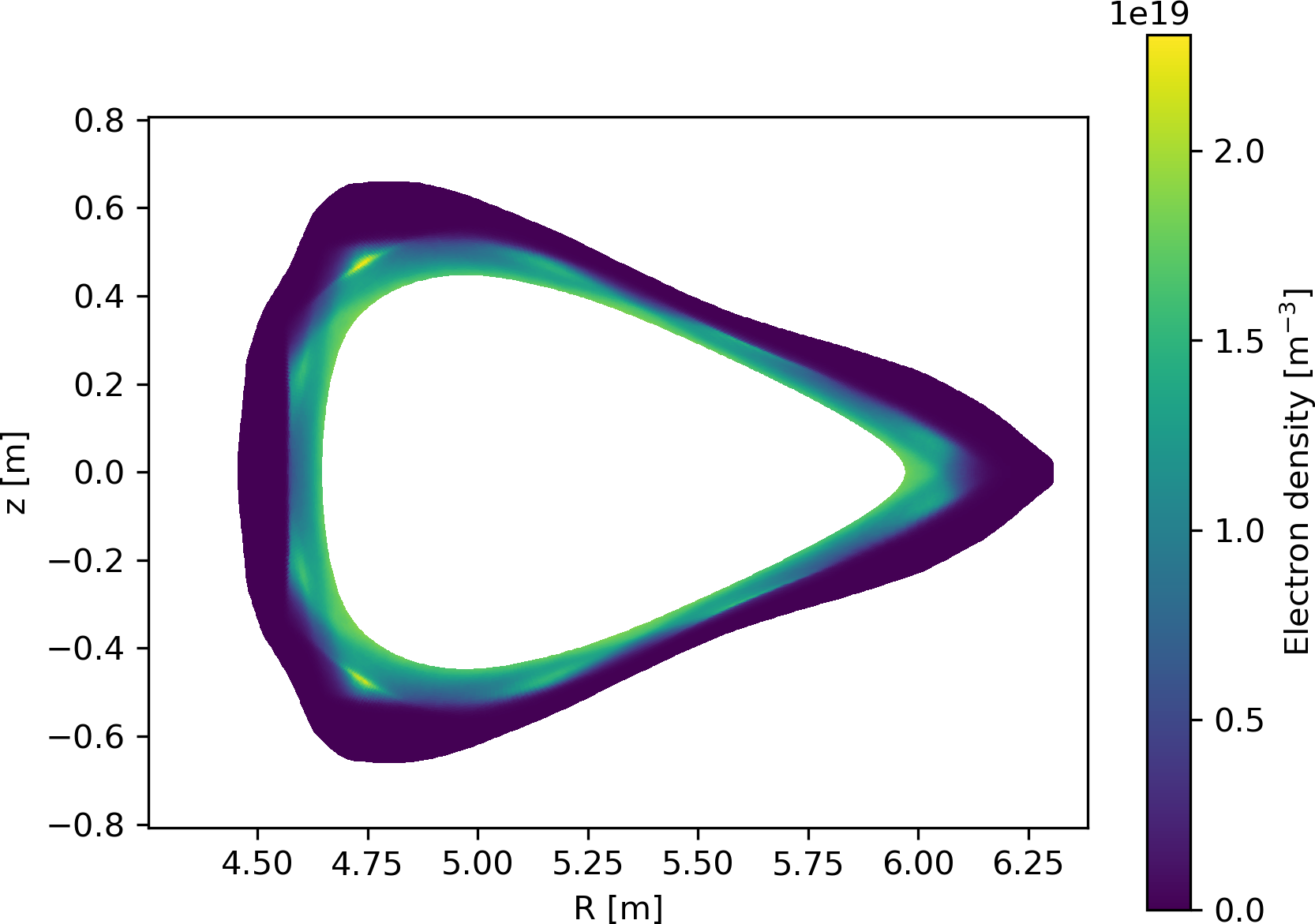}}
  \raisebox{-0.5\height}{\includegraphics[width=.4\linewidth]{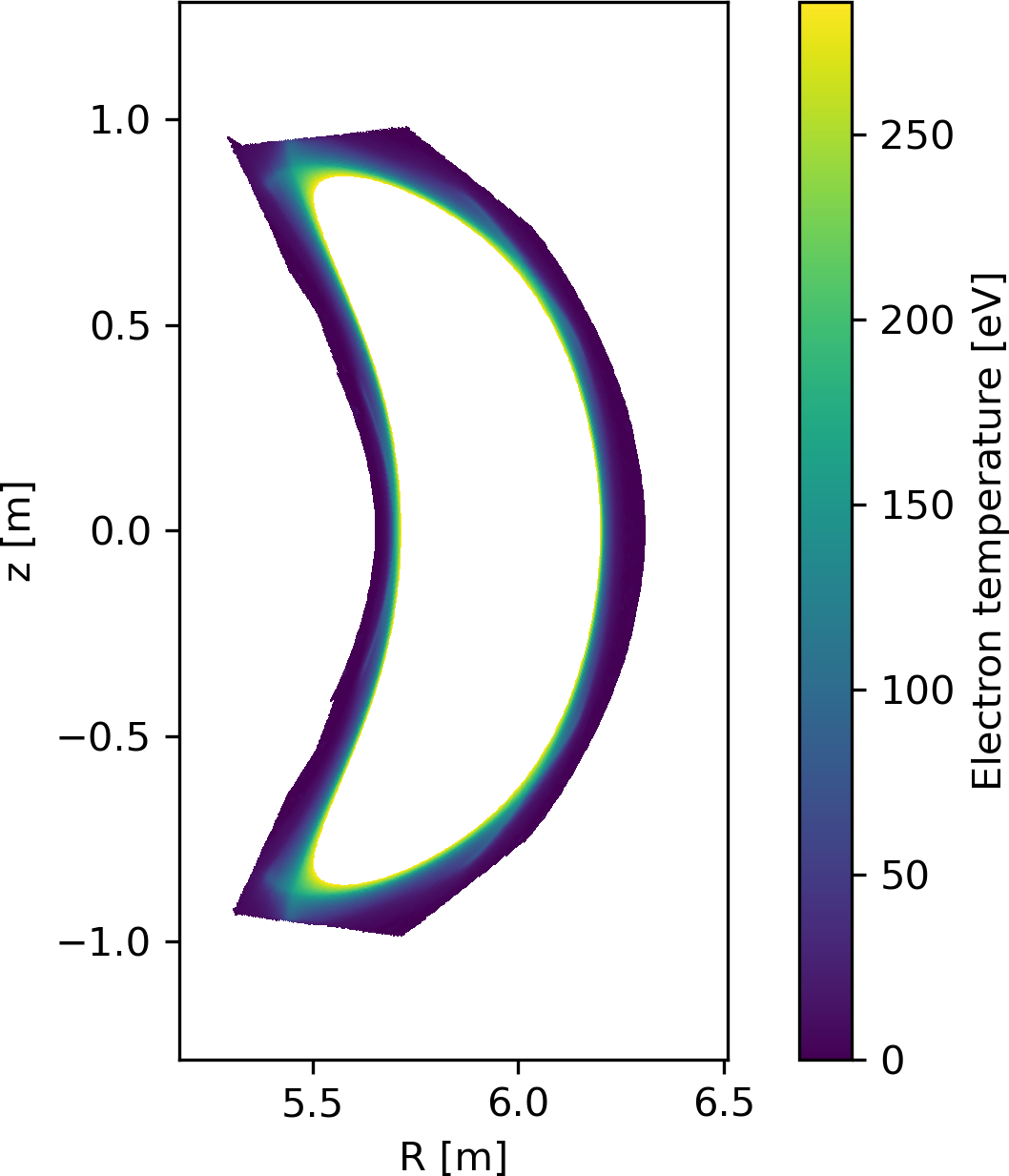}}
  \raisebox{-0.5\height}{\includegraphics[width=.5\linewidth]{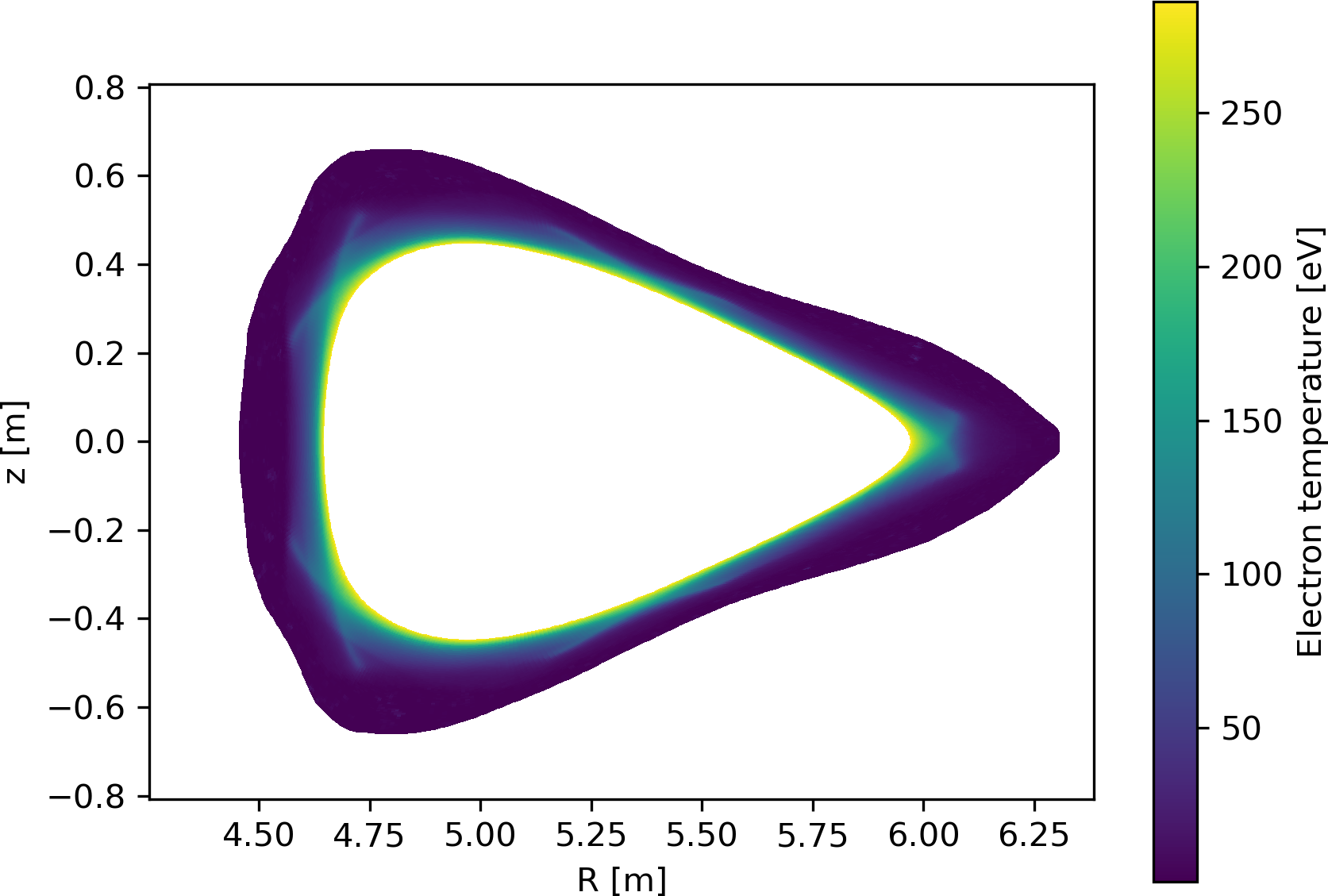}}
  \caption{Density (top) and electron temperature (bottom) profiles for the
    \dens{1} case with \Dis{0.2} at the bean shape, at $\phi=0$ (left) and the
    triangular shape, at $\phi=\pi / 5 = 36$\,\textdegree{} (right).
  }\label{f:simnt}
\end{figure}

Fig.~\ref{f:simnt} shows plots of the electron density and temperature
distribution of a simulation, where the diffusion coefficient was
set to \Dis{0.2} and the upstream density was set to
\dens{1}.  The density shows a peak just in front of the
target, at toroidal angle $\phi=0$ at the upper and lower target plates. At the triangular shape
($\phi=\pi / 5 = 36$\,\textdegree) no target plates are present and thus also the density
is not strongly peaked in the SOL.
The temperature drops of towards the target. While in this case the electron
temperature at the separatrix is around 160\,eV, the separatrix electron
temperature is in all cases below 200\,eV.
Experimentally, separatrix electron temperatures were generally
between 30 and 100\,eV.

Similar to the experimental result, shown in fig.~\ref{f:con} (right),
the main heat-flux is on the low iota target, with a strike line width and
location similar to the experimental one. The main difference is that the main
power is at \phiis{-5}, while in the experimental
figure the main heat-flux is at \phiis{-15}.

\subsection{Toroidal distribution}
\begin{figure}
  \centering
  \includegraphics[width=.99\linewidth]{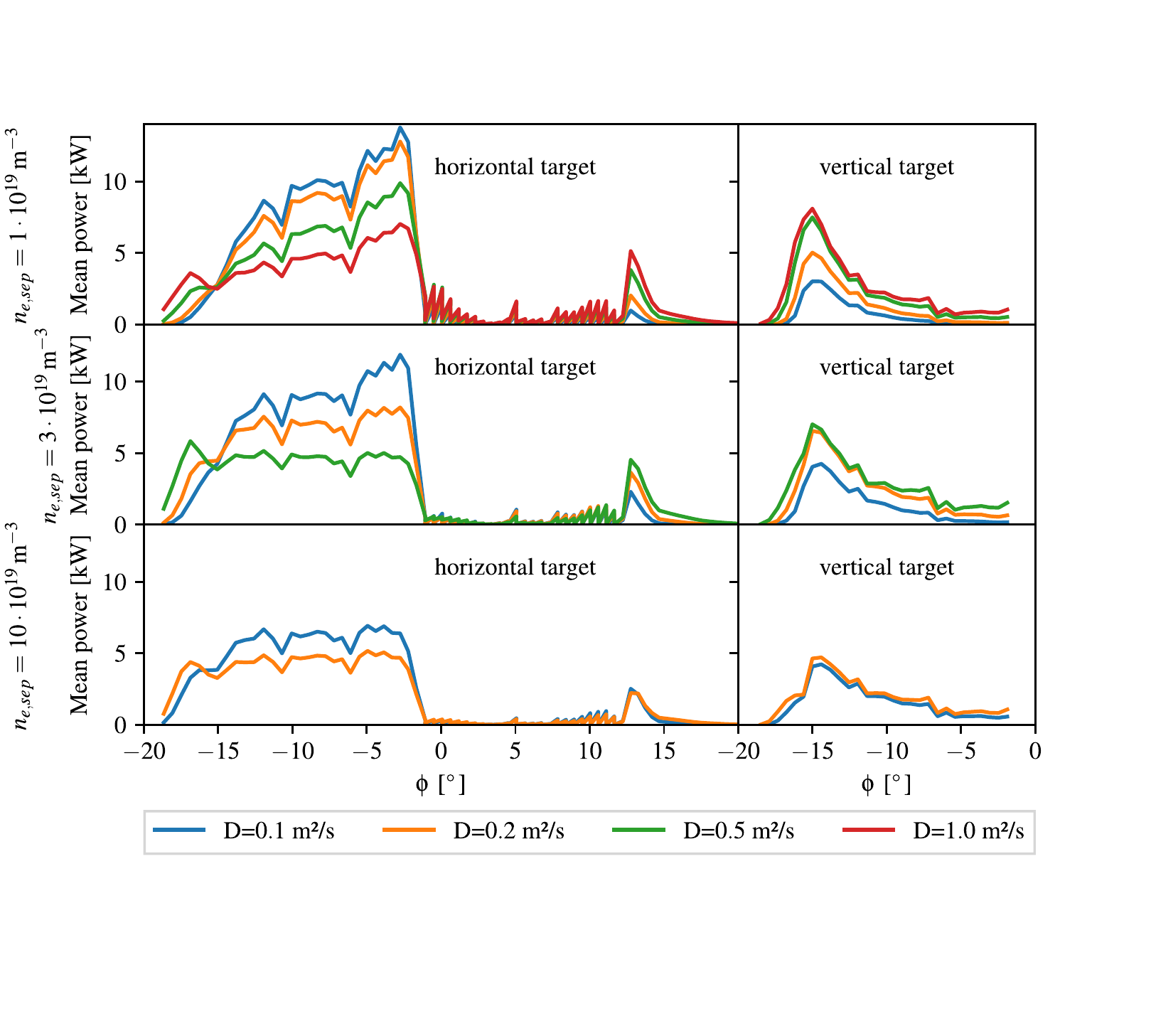}
  \caption{Plot of the power per finger for simulations with different
    separatrix density and different diffusion coefficient. This plot extends
    the density scan in fig.~\ref{f:powerfinger} with a scan of the diffusion
    coefficient $D$, with the heat diffusivity $\chi = 3D$ and $q_\perp
    \propto n \chi$.
  }\label{f:powerfingerconst}
\end{figure}
The toroidal distribution of the heat-flux, and how the change of the
diffusion parameter impacts it is shown in
fig.~\ref{f:powerfingerconst}.
\todo{@flr: I chose to start at 0.1 rather then 0.2 so that I can have
  a significant increase and still match the narrow feature. Maybe for
  scen. C 0.15 or so might have been better. 0.2 would have only
  allowed to tune until the point where it has an impact on the
  target, which is imho not as nice as restricting to a single value.
  Might be nice to add something on this - but where?}

\subsection{Strike-line}
\begin{figure}
  \centering
  \includegraphics[width=.49\linewidth]{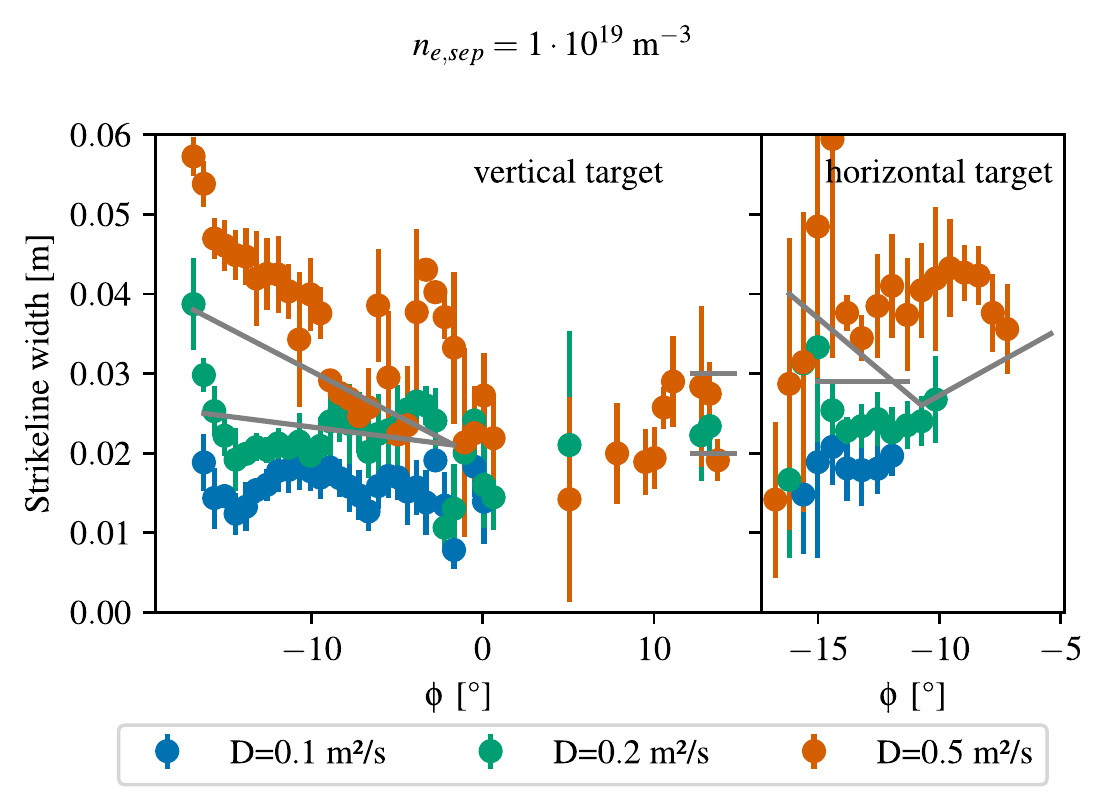}
  \caption{Mean of the strike-line width for the different fingers, as
    introduced in fig.~\ref{f:fingermap}. Note that the error bars are
    expected to be smaller than in the experimental data, shown in
    fig.~\ref{f:slw:09}, as the experimental data includes variations in time
    and across the different modules. Like fig.~\ref{f:slw:09} only fingers
    receiving at least 1\,kW of power are included.
    Simulation results for diffusion values \Dis{0.1 \ldots 0.5}
    and \dens{1}.
    The grey lines show the estimates for the low density case from
    fig.~\ref{f:slw:09}.
  }\label{f:slw:const}
\end{figure}
\begin{figure}
  \centering
  \includegraphics[width=.49\linewidth]{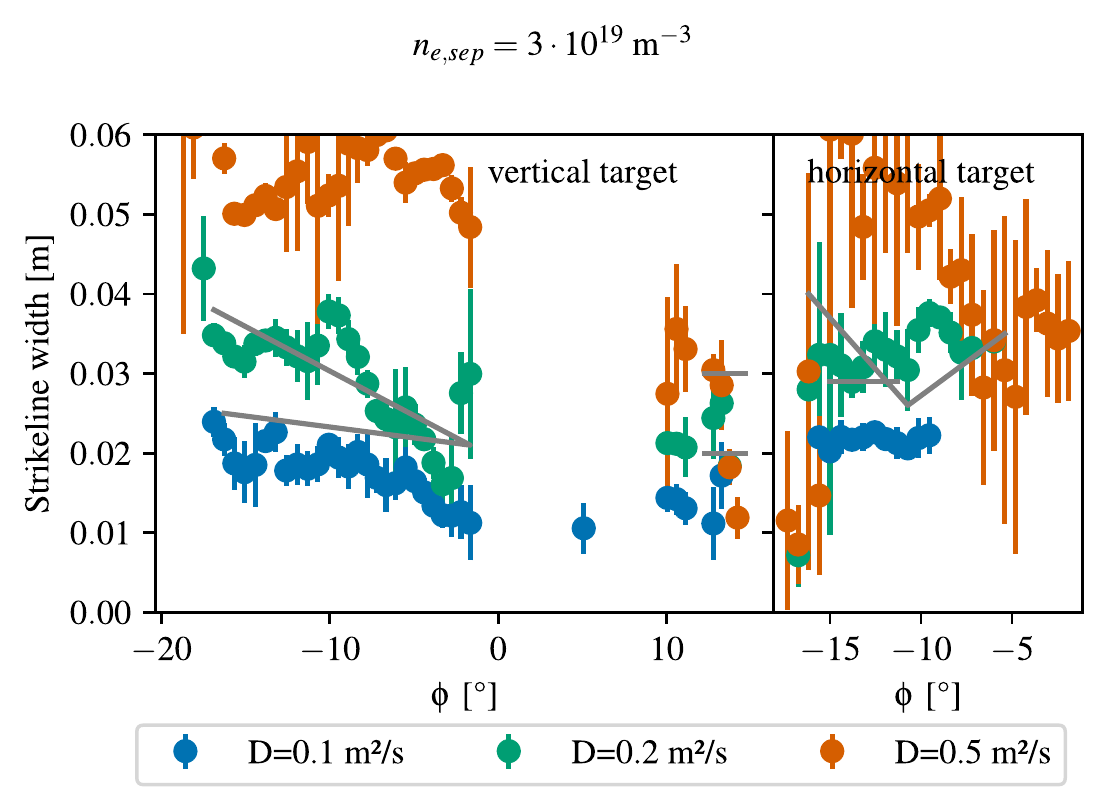}
  \includegraphics[width=.49\linewidth]{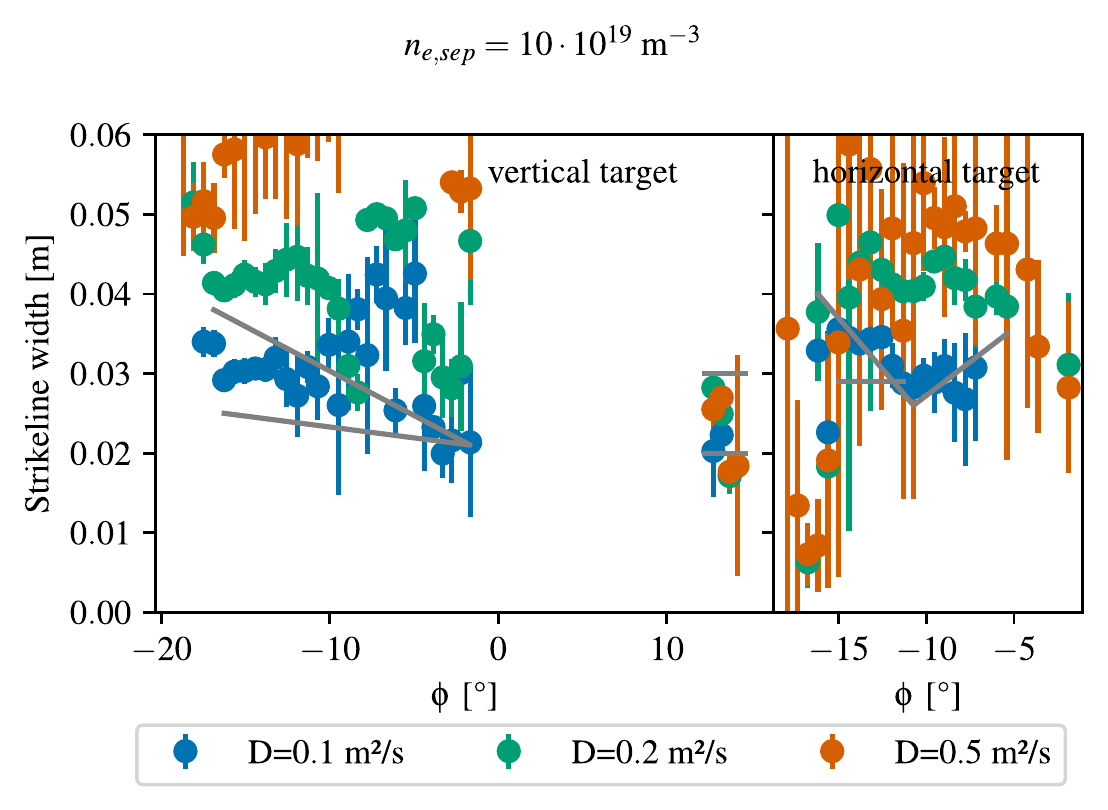}
  \caption{Mean of the strike-line width as a function of the fingers. Like
    fig.~\ref{f:slw:const} but for \dens{3} and \dens{10}.
  }\label{f:slw:const:dens}
\end{figure}
Besides the question where the power is deposited, the width of the strike
line is of interest, as that influences over what area the heat is
distributed, and thus also the peak heat-flux that the divertor has to
withstand. Besides this more practical question, the strike line width gives
also insight into the transport. Fig.~\ref{f:slw:const} shows the fitted
strike line width for a separatrix density \dens{1} with a diffusion
coefficient scan in the range
\Dis{0.1\ldots 0.5}. As mentioned, $\chi$ is scaled as $3D$.  The strike
line width for the other two density cases of interest \dens{3} and \dens{10} are shown in
fig.~\ref{f:slw:const:dens}.  The experimentally observed density are
likely between \dens{1} and \dens{3}, as will be later discussed based
on MPM data in sec.~\ref{s:mpm}.

For the smallest \Dis{0.1} the strike line width is $1\ldots 2$\,cm on the
low iota target, and thus smaller than the experimentally
observed ones, shown before in
fig.~\ref{f:slw:09}. For \Dis{0.2} the strike line width is $2\ldots 3$\,cm
matching most closely to the experiment, while for \Dis{0.5} the strike
line width is in the range of $2\ldots 5$\,cm and thus a bit larger than
experimentally observed.

The peak at \phiis{10}
agrees with experiment for the two larger diffusion
coefficients.
\Dis{0.2} matches most closely for the lower divertors.
A good match for the upper divertors is \Dis{0.5} as well as the
\Dis{0.2} for the \dens{3} case.

\subsection{Upstream data}\label{s:mpm}
To further compare the output of the models to experimental data,
upstream data is beneficial as it is separated from the location that
was optimised for.  As introduced, the MPM can measure the density and
temperature in the SOL, outside of the separatrix. Due to the
separation from the targets, this can further ensure a matching
transport model is chosen.

\begin{figure}
  \centering
  \includegraphics[width=.99\linewidth]{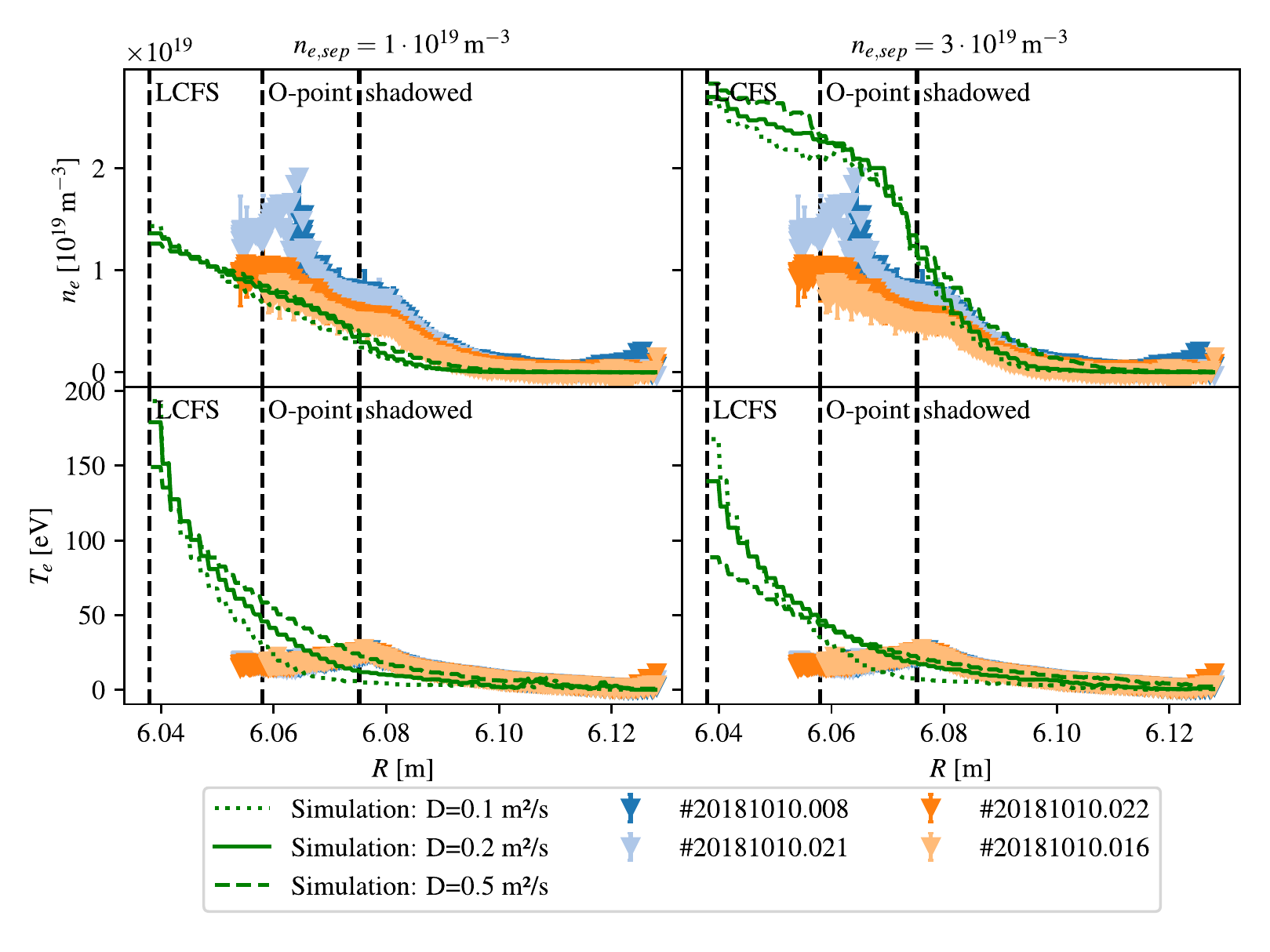}
  \caption{Plot of electron density and temperature as a function of
    the radial position. Shown is a 1D cut along the path of the MPM
    diagnostic~\cite{killer21a,killer19a}.  The experimental data from
    the MPM diagnostic is shown as symbols. The lines are the
    simulations with \dens{1} (left) and \dens{3} (right).
    Data from the \Dis{0.1} is plotted as green dotted, \Dis{0.2}
    as green line (best match based on strike line width) and \Dis{0.5}
    as green dashed.
    The point magnetically closest to the O-point, as well as the position of
    the last closed flux surface (LCFS) and the onset of the shadowed area
    featuring short connection lengths, are also plotted.
  }
  \label{f:mpm}
\end{figure}

Figure~\ref{f:mpm} shows the density and temperature of a line of sight
measured by the MPM diagnostic.
Although no experimental MPM data is is available for program
\#20180920.009, \#20180920.013 and \#20180920.017, similar programs
with MPM data exist and are used as an upstream comparison to
simulation.
These programs have similar heating power, the same magnetic configuration and
line integrated densities are in the range \lined{4\ldots 6}.
The simulation results, are plotted as lines in
fig.~\ref{f:mpm}.
The simulations matching best the strike line width are plotted as
lines, smaller $D$ distributions are dotted and larger ones are
dashed.

All simulations show essentially monotonic behaviour in the
temperature and the density. This is in contrast to the experimental
data. The experimental density shows for some cases a peak around the O-point,
while for other cases it is monotonic.
The temperature profile shows a clearly monotonic trend in the shadowed
region, in contrast to the data in the longer connection length, where a
hollow temperature profile is observed, featuring a reduction towards the
O-point.  The hollow temperature is observed in none of the simulations.
In the shadowed region a roughly exponential behaviour for $n_e$ and $T_e$ is
observed.

The separatrix density is an input parameter for the simulations. As
such we can freely choose a density, to best match the experimental
results.  For all diffusion cases, the best match is between \dens{1}
and \dens{3}.  Quantifying which case matches best, or which density
would match best, is not well defined as the profiles do not match.
\dens{3} matches very well in the shadowed area,
while, depending on the experimental measurement, the \dens{1} case
matches around the O-point.

For the temperature, all simulations feature a to high temperature
towards the island centre. The MPM was not able to measure closer to the
separatrix, but the separatrix temperature is generally estimated to be below
100\,eV.
In general with increasing $D$ the separatrix temperature decreases,
and the temperature fall-off-length increases.



\section{Discussion}\label{s:discussion}
The experimental heat-flux data from W7-X has been analysed
determining the toroidal distribution as well as using fits of the
strike line. The fits allowed to determine the position and the width of the
strike line, with the exception of the higher density case \#20180920.017
where the fit did not converge reliably, as smaller structures where
present.

Determining the shape of the strike line allows to determine how much of the
power is deposited on the main strike line versus how much in total is
deposited on the divertor.
A significant amount of power is observed by the IR cameras to be
deposited outside of the main strike-line ($\sim1\,$MW). This power is seen
as a broad feature and its cause is not yet known.
Surface layers, which have been observed to build up on certain areas
of the divertor over the campaign~\cite{mayer20a}, may increase the IR
emissivity of the targets, thus possibly creating an artefact of
higher heat flux.
Another possible explanation is power load by plasma
radiation close to the divertor.  Initial calculation seem to agree
with the radiation
hypothesis, however an in-depth study is outside of the scope of this paper.
Future work is planned to investigate this feature.

The strike line width from the experiments has been determined to be around 2
to 4$\,$cm in the magnetic standard configuration. This is true for all areas
on the target where a significant heat-flux is observed, including the low
iota target. The 2 to 4\,cm is observed independent
of the connection length.  This observation is reproduced in the simulations,
where for a given density and diffusion the strike line width is roughly
constant for all significantly loaded areas.


For the experimental data no clear trend of the strike line width with
density is observed. For the simulations a clear trend of increasing
strike line width with increasing density is observed. In the
simulations the density variation was however much larger.

While it is possible to match the strike line width to the experimental ones,
none of the simulations matched the toroidal distribution, especially at the
low iota target.  

\todo{Can this lead to transport into the far low iota?}
\todo{
Future work
~~~~~~~~~~
* Implement drifts
* Understand transport of scenario B
* rev field experiments
}

\section{Conclusion and Summary}\label{s:conclusion}
For comparing experimental and simulation data of the target heat-fluxes, a
fitting routine has been implemented, which works reliably for
well-behaved IR heat-flux data. Future work will make the fitting more
robust to counteract any possible artefacts in the heat-flux
analysis.

In the past some qualitative comparison between experiments and simulations
have been attempted for the SOL of W7-X. The here performed quantitative
comparison shows that, in order to reproduce the experimentally observed strike
line width of the range 2 to 4\,cm, diffusion coefficients in the range of
0.2\,m$^2/$s are needed in the magnetic standard configuration for
low to medium density cases.


There are significant differences between simulations and
experiments, that could not be reproduced by the simulation.
Some differences are expected to be due to a lack of drifts in
EMC3-EIRENE: e.g. the
up-down asymmetry on the divertor target plates. Additionally, small
toroidal asymmetries are observed in experiment, which are expected
due to the
non-perfect error field correction. Due to the inherent symmetry of
the simulation, they are, like drifts, not expected to be reproduced.
However, other discrepancies
remain which are not expected, such as the difference in
the toroidal distribution of the heat flux on the low iota target.
Additionally hollow temperature profiles in the islands are measured by the MPM
diagnostic, which has not been reproduced by the simulations.

Non-isotropic transport coefficients could be used in EMC3-Eirene to match
both upstream values as well as target heat-flux data.


\section{Acknowledgement}
This work has been carried out using the xarray
framework~\cite{hoyer17a,xarray_0_17_0}.
Some task have been paralellised using GNU parallel~\cite{tange18a}.

The simulation presented here are available~\cite{bold21a}

\todo{@flr: So really no funding acknowledment is needed? What about
  co-authors? Aren't at least some of them payed by eurofusion?}

\bibliographystyle{iaea_misc}
\bibliography{phd}

\end{document}